\documentclass[a4paper,fleqn,usenatbib]{mnras}
\pdfoutput=1

\usepackage{newtxtext,newtxmath}
\usepackage[T1]{fontenc}
\usepackage{ae,aecompl}

\usepackage{graphicx}	% Including figure files
\usepackage{amsmath}	% Advanced maths commands
\usepackage{amssymb}	% Extra maths symbols
\usepackage{tablefootnote}

\title[The Lack of Correlation between O32 \& $f_{esc}$]{On the Lack of
  Correlation Between [OIII]/[OII] and Lyman-Continuum Escape Fraction}

\author[R. Bassett et al.]{
R. Bassett$^{1,2}$\thanks{E-mail: rbassett@swin.edu.au (RB)},
E. V. Ryan-Weber$^{1,2}$,
J. Cooke$^{1}$,
C. G. Diaz$^{3,4,5}$,
T. Nanayakkara$^{6}$,\newauthor
T.-T. Yuan$^{1,2}$,
L. R. Spitler$^{7,8,9}$,
U. Me\v{s}tri\'{c}$^{1,2}$,
T. Garel$^{10}$,
M. Sawicki$^{11}$,
S. Gwyn$^{12}$,\newauthor
A. Golob$^{11}$
\\
$^{1}$Centre for Astrophysics and Supercomputing, Swinburne University
of Technology, PO Box 218, Hawthorn VIC 3122, Australia\\
$^{2}$ARC Centre of Excellence for All Sky Astrophysics in 3 Dimensions (ASTRO 3D), Australia\\
$^{3}$Gemini Observatory, Southern Operations Center, La Serena,
Chile\\
$^{4}$Instituto de Ciencias Astron\'omicas, de la Tierra y del Espacio
(ICATE), San Juan, Argentina\\
$^{5}$Consejo de Investigaciones Cient\'ificas y T\'ecnicas (CONICET), CABA, Argentina\\
$^{6}$Leiden Observaroty, Leiden University, NL-2300 RA Leiden, The Netherlands\\
$^{7}$Research Centre for Astronomy, Astrophysics \& Astrophotonics,
Macquarie University, Sydney, NSW 2109, Australia\\
$^{8}$Department of Physics \& Astronomy, Macquarie University,
Sydney, NSW 2109, Australia\\
$^{9}$Australian Astronomical Observatories, 105 Delhi Rd., Sydney NSW
2113, Australia\\
$^{10}$Centre de Recherche Astrophysique de Lyon, Universit\'{e} Lyon
1, CNRS, Observatoire de Lyon; 9 avenue Charles Andr\'{e}, F-69561
Saint-Genis Laval Cedex, Franc\\
$^{11}$Saint Mary's University, Department of Astronomy \&
Astrophysics and the Institute for Computational Astrophysics,
Halifax, Canada\\
$^{12}$NRC-Hertzberg, 5071 West Saanich Road, Victoria, British
Columbia, V9E 2E7, Canada
}

\date{Accepted XXX. Received YYY; in original form ZZZ}

\pubyear{2017}

\begin{document}
\label{firstpage}
\pagerange{\pageref{firstpage}--\pageref{lastpage}}
\maketitle

\begin{abstract}
We present the first results of our pilot study of 8 photometrically selected
Lyman continuum (LyC) emitting galaxy candidates from the COSMOS field and focus on their
optical emission line ratios. Observations were performed in the H and
K bands using the Multi-Object Spectrometer for Infra-Red Exploration
(MOSFIRE) instrument at the Keck Observatory, targeting the [OII],
H$\beta$, and [OIII] emission lines. We find that photometrically selected LyC
emitting galaxy candidates
have high ionization parameters, based on their high [OIII]/[OII] ratios (O32),
with an average ratio for our sample of 2.5$\pm$0.2. Preliminary results of
our companion Low Resolution Imaging Spectrometer (LRIS) observations,
targeting LyC and Ly$\alpha$, show that those galaxies with the
largest O32 are typically found to also be Ly$\alpha$
emitters. High O32 galaxies are also found to have tentative non-zero
LyC escape fractions ($f_{esc}(LyC)$) based on $u$ band photometric
detections. These results are 
consistent with samples of highly ionized galaxies, including
confirmed LyC emitting galaxies from the literature. We also perform a
detailed comparison between the observed emission line ratios and
simulated line ratios from density bounded H\,{\sc ii} regions modeled using
the photoionization code MAPPINGS V. Estimates of $f_{esc}(LyC)$ for
our sample fall in the range from 0.0-0.23 and suggest possible tension with
published correlations between O32 and $f_{esc}(LyC)$, adding weight
to dichotomy of arguments in the literature. We highlight
the possible effects of clumpy geometry and mergers that may account for such
tension.
\end{abstract}


%%%%%%%%%%%%%%%%%%%%%%%%%%%%%%%%%%%%%%%%%%%%%%%%%%

%%%%%%%%%%%%%%%%% BODY OF PAPER %%%%%%%%%%%%%%%%%%

\section{Introduction}

The epoch of reionization (EoR) is a fundamental cosmological event
marking a phase shift of the intergalactic medium (IGM) from neutral to
ionized. In the past few decades a major effort has been put in by the
astronomical community into understanding what drives this shift and
how the process of reionization proceeds. Although fairly tight
constraints have been placed on the redshift marking the end of the
EoR around $z=6$ \citep{fan06,komatsu11,zahn12,becker15}, important open
questions remain. In particular, the identity
and precise nature of the sources of the required ionizing radiation
are uncertain. 

It appears likely that the main source of ionizing photons must be
hosted by galaxies:
either massive stars in rapidly star-forming regions, or 
quasars (QSOs). There is mounting
evidence for a rapid drop in the
number density of QSOs above $z$ = 6 suggesting that these objects do
not represent the main drivers of reionization
\citep{hopkins07,jiang08,fontanot12,masters12,mitra13,ueda14,mcgreer17,hassan18}. This
observation has compelled researchers studying the EoR to focus instead on young,
star-forming galaxies 
\noindent \citep{ouchi09,wise09,yajima11,bouwens15,paardekooper15}. In such
galaxies, high
energy stars (O/B stars, Wolf-Rayet stars, and/or x-ray binaries) represent the
primary sources of ionizing photons \citep{rauw15,eldridge17}. One crucial
unknown regarding star-forming galaxies during the EoR, however, is:
what are the conditions that allow high energy photons to escape
from galaxies and subsequently ionize the IGM?

A key quantity in this framework is the fraction of ionizing, Lyman continuum
photons (LyC, $\lambda$ < 912{\AA}) that escape from star-forming galaxies
into the IGM, $f_{esc}(LyC)$. Observation of LyC photons is 
difficult, however, due to a combination of factors both internal and
external to the source galaxy. Studies of galaxies during
the EoR
predict that an average $f_{esc}(LyC)$ of $\sim$0.10-0.20 is required to
drive reionization
\citep{ouchi09,raicevic11,fontanot12,robertson13,dressler15}. Although 
hydrodynamical simulations show that LyC escape fraction can reach up to 1.0 in some
cases \citep[though only briefly, e.g.][]{trebitsch17,rosdahl18}, from observations the majority of
confirmed Lyman continuum emitting galaxies (LCEs) have $f_{esc}(LyC)$ estimates below 0.15
\citep{leitet13,borthakur14,izotov16,leitherer16}. There are a few
examples of galaxies with estimated $f_{esc}(LyC)$ as high as
0.45-0.73
\citep{izotov18,izotov18b,bian17,vanzella17,debarros16,shapley16,vanzella16,fletcher18},
however these galaxies are extremely rare. This paucity of examples of
high $f_{esc}(LyC)$ galaxies is consistent with the brevity of a high
$f_{esc}(LyC)$ phase seen in simulations, however a lack of large
observational samples makes quantifying the relative contribution from
star-forming galaxies to reionization challenging.

A related issue, and another contributor to the scarcity of observed
strong LCEs, is
the opacity of the IGM to ionizing photons. Neutral IGM gas can
attenuate, or even absorb completely, LyC photons that escape from
galaxies. 
Prior to and during the EoR, the IGM is mostly neutral and opaque to ionizing
photons \citep{inoue14,grazian16}. Coupled with the 
faintness of high redshift galaxies, this means that escaping LyC photons
from the galaxies that actually drive reionization (i.e. galaxies at $z\gtrsim6.0$) are highly unlikely to be
observed. Thus, quantifying $f_{esc}(LyC)$ during the EoR will
require a proxy for LyC escape that is more readily observed
during this epoch. In this vein, on-going projects focus
on identifying large samples of post-EoR LCEs and targeting these
galaxies at wavelengths longer than LyC in search of other galaxy
properties that are correlated with $f_{esc}(LyC)$
\citep[e.g.][]{izotov16,izotov18}. 

An example of a proxy for $f_{esc}(LyC)$ that has been under recent scrutiny is the
ratio of [OIII] ($\lambda$5007 {\AA}) to [OII] ($\lambda \lambda$3272
{\AA})\citep[the O32 ratio, e.g.][]{nakajima14}. From
photoionization modeling it has been shown that, in H\,{\sc ii} regions with
ionization bounded conditions (i.e. $f_{esc}(LyC)=0$), a value of O32 $\gtrsim$ 1
is related to high ionization \citep[i.e. a ``hard'' ionizing
spectrum,][]{kewley02,martin-manjon10}. Alternatively, O32 will also
increase in density bounded H\,{\sc ii} regions or in clumpy star-forming
regions where 
channels for LyC escape represent ``holes'' into the hot, higher
ionization, [OIII] ($\lambda$5007 {\AA}) emitting inner regions
of nebulae that bypass
cooler, low ionization, [OII] ($\lambda\lambda$3727 {\AA}) emitting
outer regions 
\citep{giammanco05,pellegrini12,jaskot13,zackrisson13,zackrisson16}. In such a
scenario, galaxies with larger O32 ratios would contain H\,{\sc ii} regions having
a larger fraction of their surfaces exhibiting conditions susceptible
to LyC escape. 

Indeed, \citet{izotov18} has
shown that low redshift LCEs are found to have O32 ratios $>$ 5, and that O32
appears to correlate with $f_{esc}(LyC)$. Furthermore, Ion2 \citep{debarros16}, the one example
of a high redshift LCE also having measured fluxes of [OIII]
($\lambda$5007 {\AA}) and
[OII] ($\lambda\lambda$3727 {\AA}), has an $f_{esc}(LyC)$ and O32 similar
to the strongest low redshift LCE from \citet{izotov18}. Whether or
not the O32 ratio has a direct correlation with $f_{esc}(LyC)$ is
still a matter of debate, although recent results refute such a strong
relationship \citep{izotov18b,naidu18}. What remains
to be done is to greatly increase the samples of 
LCE at both low and high redshift in order to conclusively test
the relationship between $f_{esc}(LyC)$ and the O32
ratio.

In this paper we present the optical line ratios, including O32, for a
pilot sample of galaxies selected as LCE candidates based on
photometric criteria following the work of
\citet{cooke14}. A majority of confirmed LCEs in the literature to date
have been selected for observations of LyC either based on already known
spectral properties  \citep[e.g. spectroscopic redshift or O32
][]{vanzella15,izotov16,izotov18} or
involve targeting lensed galaxies
\citep[e.g.][]{bian17,vanzella17} or narrow-band selected Ly$\alpha$
emitters \citep[LAEs,][]{fletcher18}. Our approach, based on a
purely photometric selection of candidate LCEs, represents a
more efficient method of identifying a large sample of LCEs as we are not required to
already have spectroscopic observations in hand or to focus
on limited samples of lensed objects. The main driver of our method is
to measure the total population of LCEs and study the full range of
galaxy properties. We note, however, that this method
also requires an accurate estimate of the photometric redshift, which
currently limits us to well surveyed fields with dense, multi-filter
photometry. 

This paper is organized as follows: in Section \ref{section:sampsel}
we describe our photometric selection methodology and present our pilot
sample of LCE candidates, in Section \ref{section:obsdr} we describe
the observations and data reduction for our selected sample, in
Section \ref{section:analysis} we describe our analysis including both
spectral measurement and photoionization modeling of optical line
ratios, in Section \ref{section:results} we present the results of
these analyses, in Section \ref{section:discussion} we discuss the
implication of our results and evaluate the success of our selection
methodology, and finally, in Section \ref{section:snc} we summarise
our study and list our conclusions.

\section{Sample Selection}\label{section:sampsel}

In this Section we describe the selection methodology of
this pilot survey. 
The goal of our photometric selection approach is to identify star-forming
galaxies at $z\gtrsim3$ that are likely to be leaking LyC photons into
the IGM. This selection uses 30-band photometry from the Cosmic
Evolution Survey \citep[COSMOS,][]{scoville07} field, including
FourStar Galaxy Evolution Survey \citep[ZFOURGE][]{straatman16} deep
medium-band imaging. The depth and wavelength coverage
of this photometric dataset provides photometric redshifts with
accuracies of $\Delta z<$2\% due to deep medium-band filter
observations probing the Balmer break. Using galaxies from the ZFOURGE
survey enables us to reliably select galaxies at $z>3.0$. 

LyC emitting galaxies at
$z>3.0$ will have some contribution from LyC photons in the
$u$ band, making this the critical observation band for our
selection. This is demonstrated in Figure \ref{fig:lcgspec} where we
show stacked Lyman Break Galaxy (LBG) spectra from \citet{shapley03} where the top and
bottom quartiles in Ly$\alpha$ equivalent width (EW) are shown in blue
and red, respectively. The lower x-axis indicates rest-frame
wavelengths while the upper x-axis shows the observed wavelengths at
the average redshift of our final sample, $\langle z \rangle=3.17$. We have
overplotted the Canada France Hawaii Telescope (CFHT) $uS$ and
$r'$ band transmission curves for comparison, indicating the spectral
regions probed by these bands for LBGs at $z=3.17$. We have also
plotted the transmission curve of the CFHT Large-Area U-band Deep Survey (CLAUDS, Sawicki et
al. in preparation) $u$ band with a cyan
dashed line. This filter is essential for estimating the level of LyC
escape in our galaxy sample as described in Section
\ref{section:fescphot}, however we note that this filter was not
available when our sample was selected. At this redshift,
observed flux in the $uS$ band is made up of contributions from both
LyC and Ly$\alpha$ forest photons.

In Figure \ref{fig:lcgspec} we have also added 
LyC flux to the LBG spectra below $\lambda_{rest}=912$ {\AA}, constant in
$F_{\nu}(\lambda)$, representing escaping LyC flux. The level of the
added flux is chosen as a fraction of the UV continuum flux, $F_{\nu}(UV)$, defined
as the median flux in the wavelength range 1450 {\AA} $<$ $\lambda$
$<$ 1550 {\AA}. The two values of $F_{\nu}(LyC)/F_{\nu}(UV)$ of 0.24 and 0.012
shown in Figure \ref{fig:lcgspec} correspond to $f_{esc,rel}(LyC)=1.0$ and
$f_{esc,rel}(LyC)=0.05$, respectively, assuming an intrinsic LyC to UV
flux ratio of 0.33 \citep[e.g.][]{vanzella12} and an IGM attenuation at
912 {\AA} of 0.72. This latter assumption is taken as the average
value of the $z=3.17$ IGM attenuation at 912 {\AA} computed following the
models presented by \citet{inoue14}. We must clarify that here $f_{esc,rel}$ is the
relative escape fraction, and a dust correction must be applied to the
observed $F_{\nu}(UV)$ to assess the absolute escape fraction (see
Section \ref{section:fescphot} for complete definitions of
$f_{esc,rel}(LyC)$ and $f_{esc}(LyC)$).

Galaxies selected using the criteria described here are part of a
pilot program to obtain optical emission lines using the
Multi-Object Spectrometer for Infra-Red Exploration (MOSFIRE)
instrument (this paper) and to detect LyC emission using 
the Low Resolution Imaging Spectrometer
\citep[LRIS,][]{oke98,steidel04} instrument (Me\v{s}tri\'{c} et al. in preparation). 

\subsection{Photometric Galaxy Selection}\label{section:colsel}

\begin{figure}
  \includegraphics[width=\columnwidth]{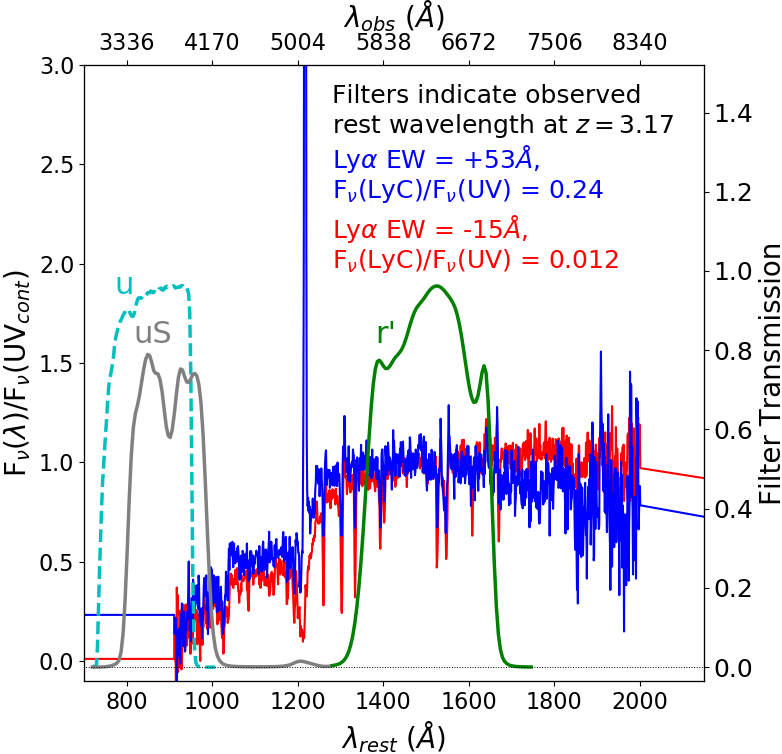}
  \caption{Composite Lyman break galaxy (LBG) spectra from \citet{shapley03}
    demonstrating to contribution from LyC photons in the $u$
    band. In blue and red we show the stacked spectra for the 
    highest and lowest quartiles in Ly$\alpha$ EW from LBGs in
    \citet{shapley03}, respectively. Raw transmission curves for CFHTLS
    $uS$ and $r'$ filters are shown with 
  wavelength coverage indicating the observed restframe wavelengths for
  galaxies at a redshift of $z=3.17$ (observed wavelengths are shown
  on the top x-axis). At $z=3.17$, the $r'$ band probes the restframe
  UV continuum around 1500 {\AA}, a wavelength regime commonly used in
LyC escape studies \citep[e.g.][]{steidel18}. We also show the CLAUDS
$u$ band, used to estimate $f_{esc}(LyC)$ in Section
\ref{section:fescphot}, which has many advantages over the $uS$ band
in regards to studying $z\sim3$ LCEs (see Section \ref{section:caveats}).}
  \label{fig:lcgspec}
\end{figure}

\begin{table*}
  \begin{center}
  \caption{Basic Properties of Photometrically Selected MOSFIRE Target
  Galaxies}
  \vspace{2mm}
  \begin{tabular}{ c c c c c c c c c}
    \hline\hline
    ID & RA & DEC &\textit{z}$^{a}$ & log$_{10}(\mathcal{M}_{*})^{b}$ &
  log$_{10}$(SFR)$^{c}$ & $uS$ & $r'$ & $Ks$\\
    & deg. & deg. & & (M$_{\odot}$) & (M$_{\odot}$yr$^{-1}$) & (mag) &
                                                                       (mag)
                                      & (mag) \\
    \hline
    12676 & 150.21806 & 2.31344 & 3.12 & 9.80 & 1.92 & 26.53$\pm$0.22
                               & 24.71$\pm$0.04 
                                                    & 23.47$\pm$0.08\\
    13459 & 150.20192 & 2.32179 & 3.18 & 9.75 & 0.75$^{d}$ &
                                                             26.63$\pm$0.30
                               & 24.72$\pm$0.03 & 23.59$\pm$0.08\\
    14528 & 150.15553 & 2.33388 & 3.08 & 9.49 & 1.70 & 26.61$\pm$0.21
                               & 24.75$\pm$0.04 
                                                    & 23.92$\pm$0.09\\
    15332 & 150.16920 & 2.34218 & 3.21 & 9.68 & 1.70 & 26.78$\pm$0.23
                               & 24.95$\pm$0.06
                                                    & 24.10$\pm$0.10\\
    15625 & 150.13919 & 2.34531 & 3.23 & 9.77 & 1.58 & 27.29$\pm$0.36
                               & 24.82$\pm$0.04 
                                                    & 23.50$\pm$0.07\\
    16067 & 150.20010 & 2.34941 & 3.21 & 9.19 & 1.81 & 26.06$\pm$0.13
                               & 24.02$\pm$0.02 
                                                    & 22.90$\pm$0.04\\
    17251 & 150.13780 & 2.36102 & 3.12 & 10.06 & 2.09 & 25.68$\pm$0.12
                               & 23.15$\pm$0.03
                                                    & 22.86$\pm$0.05\\
    17800 & 150.17387 & 2.36797 & 3.21 & 9.43 & 1.90 & 26.28$\pm$0.19
                               & 24.26$\pm$0.03 
                                                    & 23.46$\pm$0.07\\
    \hline
  \end{tabular}\label{table:1}\\
  $^{a}$ZFOURGE photometric redshift\\
  $^{b}$Stellar Mass measured from ZFOURGE SED fitting with emission
  lines included\\
  $^{c}$ZFOURGE UV+IR SFR \citep[][IMF]{chabrier03}, with IR data coming from Spitzer/MIPS and
  Herschel/PACS observations \citep{tomczak16}\\
  $^{d}$SFR for galaxy 13459 is based on UV alone as it is undetected
  in the far IR bands
  \end{center}
\end{table*}

The goal of our sample selection for the galaxies presented in this
paper was to identify galaxies near $z\sim 3$ that
are likely to exhibit nonzero LyC escape fractions, and to select
those with LyC fluxes that could be detected
during a single night of LRIS observations.
As described in Section \ref{section:fescphot}, an increase in LyC
flux for galaxies at $z\sim3.17$ will 
produce an increase in the observed $uS$ band flux beyond what is
expected from Ly$\alpha$ forest photons alone. Thus, for our sample,
we select from those galaxies at $z\simeq3$ with the brightest $uS$
fluxes from our parent sample. 

Although selecting
galaxies at $z>3.7$, where the $uS$ band will have little or
no contribution from $\lambda>912$ {\AA} photons, will produce a
cleaner selection of LyC emitters,
we select galaxies close to $z=3.0$ for three reasons:
\begin{itemize}
\item Galaxies must be at $z\gtrsim2.5$ for LyC photons to be
detectable by ground-based spectroscopy (e.g. LRIS), however the
intrinsic faintness of LCEs along with the sensitivity of UV detectors
at very short wavelength puts an effective lower redshift limit of
$z\simeq2.9$ for ground based observations
\item From $z=4.0$ to $z=3.0$, the average transmission of the IGM
  to radiation at 912 {\AA} increases from $\sim$0.4 to $\sim$0.7 \citep{inoue14}
\item Observing galaxies at $z=3.0$ provides a gain in brightness of
  $\sim$0.5 mag compared to $z=4.0$
\end{itemize}
Given these three factors, for galaxies with the same $L(LyC)$ and
$f_{esc}(LyC)$, we expect those at $z\sim3$ to 
be the most likely to have LyC spectroscopically detected in a single
night of observations.

This selection requires a parent sample of galaxies with
accurate photometric redshifts above $z\simeq3$ to be sure that the
$uS$ band samples significantly the LyC portion of the spectrum.
For this reason, our sample is pre-selected
from the ZFOURGE survey
\citep{straatman16}. Using SED fitting to 
up to 30 photometric 
bands, ZFOURGE provides medium-band, IR, photometric
redshift accuracy of $\Delta z<$2\% up to $z=4.0$ (with a maximum
value at $z=3.0-4.0$ of $\Delta z \simeq$5\%). Furthermore, the ZFOURGE
survey footprints are in HST legacy fields, thus providing space-based
imaging to check for line of sight contaminants
\citep[e.g.][]{vanzella10,vanzella12}. Initial ZFOURGE SED
fits include the $uS$ band using templates assuming $f_{esc}(LyC)=0.0$
(i.e. LBG-like spectra), however we performed additional SED fits for
our candidate galaxies excluding the $uS$ band with no significant
change in photometric redshift. We also note
that ZFOURGE galaxies are selected
based on $Ks$ band magnitude, thus low-mass galaxies may be missing
from their catalogs \citep{spitler14}.

\begin{figure*}
  \includegraphics[width=\textwidth]{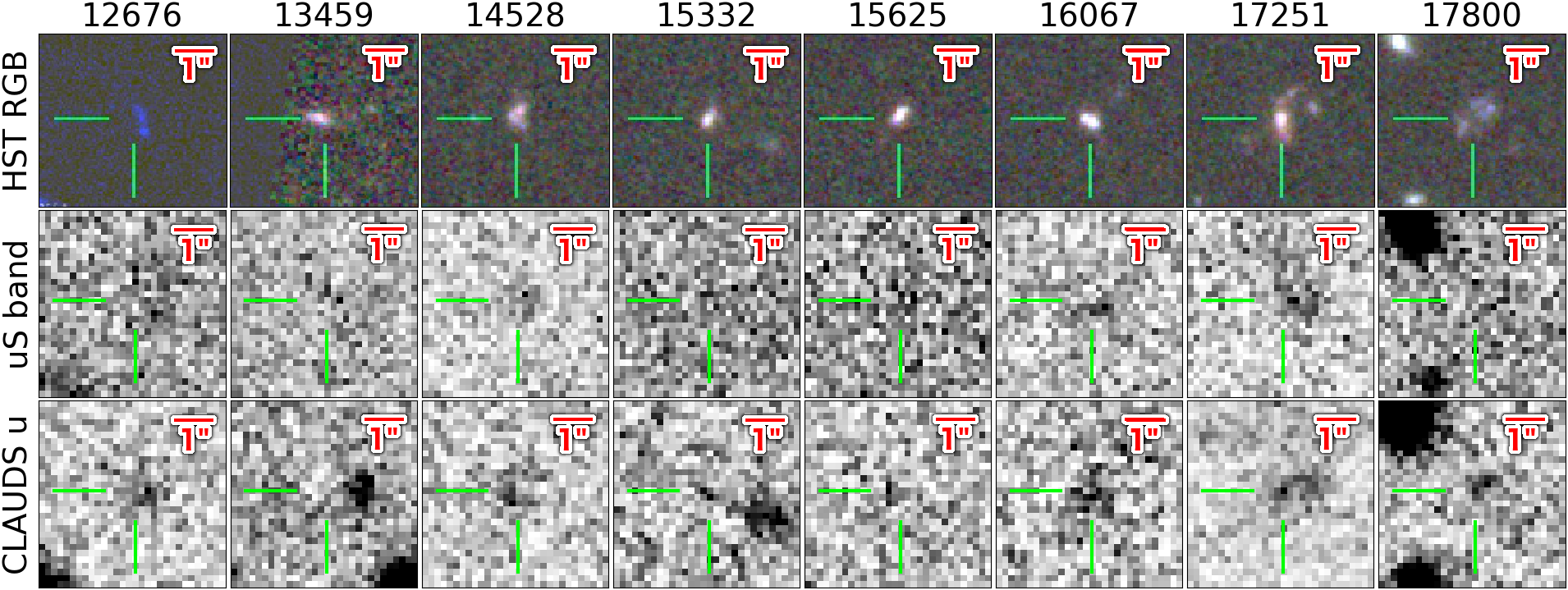}
  \caption{\textit{Top:} RGB images for our selected galaxies produced using HST
    F160W, F125W and F814W filters (galaxy 12676 only has coverage in F814W). HST imaging shows
    galaxies in our sample to
    be elongated, nonsymmetric, and/or exhibiting multiple continuum
    peaks. These features may be indicative of ongoing mergers;
    however, we cannot fully rule out the possibility of low redshift,
    line-of-sight companions. Although there is no evidence from our
    MOSFIRE spectra of such companions, this possibility will affect our
    selection, and we discuss the implications of this in Section
\ref{section:future} \textit{Middle:} CFHTLS $uS$ images of our galaxy
sample. \textit{Bottom:} CLAUDS $u$ band images of our galaxy sample
that are used in estimating $f_{esc}(LyC)$ in Section \ref{section:fescphot}.}
  \label{fig:HST}
\end{figure*}

After our ZFOURGE photometric redshift pre-selection, our main
selection criterion is that galaxies are sufficiently bright in the
$uS$ band such that their rest-frame UV spectra can be observed in a
single night of LRIS observations. We estimate, based on previous
experience, that our $uS$ detection limit for our instrumental setup is
26-26.5 mag with the evidence of fainter continuum flux using spectral
binning. In the context of LyC emitting
galaxies, those galaxies with the highest $uS$ flux at $z\simeq3$ are
the most likely to have nonzero $f_{esc}(LyC)$, assuming galaxies are
selected at a similar $Ks$ band luminosity (i.e. similar stellar
mass). Nonzero $f_{esc}(LyC)$ would manifest as an excess $uS$ flux
relative to what is expected assuming the observed $uS$ flux
originates entirely from Ly$\alpha$ forest photons (see also Section
\ref{section:fescphot}).  

Following the criteria described above, we focus our
selection on galaxies with ZFOURGE photometric redshifts of
$z\simeq3.2$ with brighter than expected $uS$ magnitudes. We are also limited on the
sky by the fact that our observations comprised of a single MOSFIRE
mask resulting in a final sample of eight galaxies. The fact that our
observations were limited to a single MOSFIRE mask also means that only 3/8 galaxies are
brighter than 26.5 mag in $uS$. The remaining galaxies were selected
as the brightest $uS$ detected galaxies falling within the footprint
of our MOSFIRE mask. Any additional unfilled slit positions were used
for ZFOURGE targets from other observational programs.

The
basic properties of our eight galaxies observed with MOSFIRE are 
described in Table \ref{table:1} and CFHT Legacy Survey (CFHTLS) $uS$ images are shown in
the middle row of Figure \ref{fig:HST}. Table \ref{table:1} shows
that 6/8 selected galaxies are well detected in the $uS$
band, having $uS$ magnitude errors smaller than m=0.25. As stated, this is
deliberate as brighter uS targets are the most likely to be detected
at LyC in a single LRIS night (or $\sim$6 hr exposure). Also included in Table
\ref{table:1} are the $r'$ band magnitudes, which corresponds roughly to the
restframe 1500 {\AA} continuum at the redshifts of our sample, for
comparison to other works. Finally, it should be noted that galaxies
in our selected sample fall within the standard LBG $(g-r)$ vs $(u-g)$
selection box \citep[e.g.][]{steidel96} and would thus be classified as LBGs.

Considering a parent sample of 717 galaxies from the
ZFOURGE COSMOS field in the redshift range 2.9 $<$ $z$ $<$ 3.2, the
three brightest targets in our sample are among the brightest 25\% in $uS$. The
remainder of our sample range from being among the brightest 31\% to
59\%. Our ongoing sample selection of LyC detected galaxies requires
careful visual inspection of images, and on average we exclude
$\sim$95-98\% of galaxies in our target redshift range. 
Typically, galaxies with $uS$ $>$ 26.5 mag are contaminated
by nearby sources (either stars or bright, low-redshift galaxies)
while sources fainter than 26.5 are very often found to be either multiple
sources in HST imaging (possible low redshift interlopers or mergers) or
undetected in $uS$. The majority of our current selections of LyC
detected galaxies fall in the range between 27.5 $<$ $uS$ $<$ 26.5,
similar to the sample presented here. The one caveat here is that, as we
describe in Section \ref{section:future}, ongoing sample selection has
shifted to $z>3.4$ as a result of now using the CLAUDS $u$ band
filter (described below). A more detailed description of our ongoing
selection of photometrically detected LyC emitting galaxies will be
presented in Cooke et al. (in preparation) and Me\v{s}tri\'{c} et
al. (in preparation).

We also estimate $f_{esc}(LyC)$
for our sample 
from $u$ band photometry using \citet{shapley03} LBG spectral stacks
(as shown in Figure \ref{fig:lcgspec}) as templates for the underlying
spectra of our sample in Section \ref{section:fescphot}. This estimate
is made using CFHT Large-Area U-band Deep Survey (CLAUDS, Sawicki et
al. in preparation) imaging,
shown in the bottom row of Figure \ref{fig:HST}, rather than CFHTLS
$uS$ for three reasons: CLAUDS imaging is deeper having a depth of
$\sim$27-28 mag (depending on position on the field),
the new $u$ filter used in CLAUDS does not suffer from red-leak
issues as described in Section \ref{section:caveats}, and it covers
bluer wavelengths having less contribution from the Ly$\alpha$ forest
(see Figure \ref{fig:lcgspec}). We note that CLAUDS imaging
was not available during the initial selection of our sample, and thus
was not included in our selection criteria.

Finally, we comment on the possibility of whether any of the eight galaxies in our
sample harbour active galactic nuclei (AGN). This issue has been
explored by the ZFOURGE team in \citet{cowley16} where galaxies are
examined in X-ray, radio, and infrared (IR) data for evidence of AGN
activity. All eight galaxies in our sample are not identified as AGN
using any these datasets. It should be noted though, that the redshift of our
sample is near the upper limit for AGN identification by these
methods. \citet{cowley16} estimate a lower limit for detecting radio
AGN of $L_{\rm 1.4GHz} = 1.9\times10^{24}$ W Hz$^{-1}$ and for X-ray
AGN of $L_{X} = 7.0\times10^{42}$ ergs s$^{-1}$ at $1.8 < z <
3.2$. Galaxies in our sample are also only weakly detected in the IR
4.5, 8.0, and 24 $\mu$m bands (or not detected in the case of 13459)
thus, from ZFOURGE data, it was not
possible to rule out the possibility of weak AGN activity in our
sample at the time of sample selection. After our spectroscopic
observations, however, our deep LRIS spectra strongly rule out AGN, which
are easily identified in the FUV through detection of high ionization
emission lines such as N \small V \normalsize ($\lambda\lambda$1240
{\AA}), O \small VI \normalsize
($\lambda\lambda$1035 {\AA}), or C \small III \normalsize
\citep[$\lambda\lambda$1907 {\AA},][]{hull12}. These are
completely absent from our observations. 

\subsubsection{Sample Selection Caveats}\label{section:caveats}

The first caveat regarding selecting only those galaxies with
strong $uS$ band detections is the danger of line-of-sight
contamination from low $z$ galaxies. This occurs where less luminous
galaxies at $z<3$ fall at a low angular separation from a target
galaxy such that the two galaxies are not well resolved in ground-based photometry. Indeed, low redshift interlopers have been found to
refute suspected LyC detections in recent studies at high
redshift \citep{vanzella10,vanzella12}. Quantitatively,
\citet{cooke14} use CFHTLS to estimate that, for $z=3-4$ galaxies down to a $u$
magnitude of $\sim$28,one can expect a $\sim$5-7\% rate of
line-of-sight contamination from lower redshift galaxies in standard
2$\farcs$0 apertures and SExtractor-based detections \citep[consistent
with previous searches for LCEs, e.g.][]{shapley06,nestor11}. 

The ZFOURGE survey includes Hubble Space Telescope (HST) photometry in
3-4 bands (depending on the location of a given galaxy), allowing us
to check for low-$z$ companions at 0.05 arcsec/pixel spatial
resolution. We show
RBG color-composite images for our sample in the top row of Figure
\ref{fig:HST}. Here the R, G, and B images come from the HST F160W, F125W,
and F814W, respectively. HST images of our sample show
a range of morphologies, including single clumps, elongation, and
multiple continuum peaks. The origin of the apparent clumpy nature of these
galaxies is unclear due to the fact that we are sampling the
restframe UV and blue optical
portion of the spectrum for star-forming galaxies
\citep{elmegreen05,forsterschreiber11,swinbank11,wisnioski12}. 
It is possible these are signatures of multiple, large star-forming regions or ongoing
mergers. We have not excluded possible mergers from this pilot survey
based on HST morphology as we are also interested in the role that
galaxy mergers play in LyC escape. The effect of mergers on optical
diagnostics is discussed further in Sections \ref{section:17251} and \ref{section:mappings2}.

The second caveat to our selection is that there is a known issue of
redleak in the CFHT $uS$
filter, which overlaps with Ly$\alpha$ at $z\sim3.17$ (see Figure
\ref{fig:lcgspec}). This means that the observed $uS$
band flux contains $\sim$1\% of the flux from redder UV continuum
and/or Ly$\alpha$. We simulate this effect for each of the 5 galaxies
in our sample that are known LAEs from our LRIS
spectroscopy with Ly$\alpha$ EW in the range
from 20-104 {\AA} (see Table \ref{table:fescphot}). This is done by replacing the Ly$\alpha$ emission
line in stacked LBG spectra from \citet{shapley03} with a Gaussian
profile having an EW matching the observed value. Here we use the
stacked LBG spectrum from \citet{shapley03} representing the top
quartile in Ly$\alpha$ EW. We then shift the observed wavelengths of
the spectrum to match the redshift of a given galaxy and measure the
relative contribution to the $uS$ band flux from photons in the
redleak portion of the transmission curve. At a given redshift, the
redleak contribution to $uS$ mag will depend on both $f_{esc}(LyC)$
and Ly$\alpha$ EW, with the maximum occurring at low $f_{esc}(LyC)$ and
high Ly$\alpha$ EW. In the case of $f_{esc}(LyC)=0.0$, we estimate a
maximum redleak contribution to the $uS$ flux of 8-17\% for LAEs in our
sample. Thus, even in the worst case, redleak will have minimal impact
on our sample selection. All future samples for this program eliminate
this issue by using deep CLAUDS $u$-band data, which does not suffer
from redleak.

The final caveat to our pilot survey sample selection is the fact that,
as we have mentioned, at $z<3.6$ the observed $uS$
magnitudes will also be influenced by photons from the Ly$\alpha$
forest  ($\lambda$ > 912 {\AA}). Clearly, this limits our ability to select
high $f_{esc}(LyC)$ galaxies based on $uS$ magnitude
alone at $z\sim3.2$. As we have noted, we select 
galaxies bright in the $uS$ band
for this pilot study to ensure spectral detection of UV 
photons with LRIS in a single night of observations, thus biasing
our sample towards lower redshifts. We note that the CLAUDS $u$ band,
which we use to estimate $f_{esc}(LyC)$ in Section
\ref{section:fescphot} has a sharp cutoff at $\sim$4000 {\AA} thus
reducing this issue. 

The results of
this pilot study inform more stringent photometric selection
criteria aimed at selecting exclusively high $f_{esc}(LyC)$
targets for future observations. Most importantly, targets are
selected at $z>3.4$ to ensure that detected $u$
band flux originates in the LyC portion of the galaxy SED. This is described
further in Section \ref{section:future} and Cooke et al. (in preparation). 

\section{Observations and Data Reduction}\label{section:obsdr}

\subsection{MOSFIRE}\label{section:mosfobs}

MOSFIRE observations for this study were performed on 01 February 2015. 
Of particular interest for our study is the line ratio of [OIII]
($\lambda$5007{\AA}) to [OII] ($\lambda \lambda$3727 {\AA}). Targeting
these lines (with the addition of H$\beta$) requires
us to observe in both the H-band and K-band (targeting [OII] and [OIII]
at $z\sim3.17$ respectively). 

For each band we estimate the average
seeing using the continuum detection of the flux calibration star
observed in a slit with a width of 3$\farcs$0. We first identify windows in the 2D
stellar spectrum that have both a relatively strong continuum
detection and lack large residuals from sky line subtraction. We
construct multiple profiles of our continuum detection along the slit
by averaging along the spectral direction in each of these windows. We
then fit these profiles using a Gaussian function. We take the final
seeing (given below for each band) as
the average value of all full width half max (FWHM) of our Gaussian fits for
each clear spectral window for a given band.

K band observations were performed using the standard MOSFIRE slit
width of 0$\farcs$7 with a resolution of $R=3610$ (velocity resolution
$\sim$40 km s$^{-1}$) and an average seeing of 0$\farcs$79. The total, on-source integration time was
3 hours using an ABAB dither sequence with the offset position
separated by 3$\farcs$0 along the slit. Individual exposures were 180
seconds with the full integration composed of three sets of 20 exposures.
For the eight galaxies in our MOSFIRE sample, we find a signal
to noise of our [OIII] ($\lambda$5007 {\AA}) line flux
ranging from 5.2 to 42.1 with an average value of 23.0. Here our noise
is defined as the integral of a Gaussian function with a central flux
density equal to the one sigma spread of our noise spectrum at the
location of a given line and a line width equal to that measured for
the line (line fitting is described in detail in Section
\ref{section:elf}). 

H band observations were performed similarly, using the same
0$\farcs$7 slit with a slightly higher resolution of R=3660 and an
average seeing of 0$\farcs$96. Given that the H-band is slightly less
sensitive than the K band, the total
on-source integration time was increased to 3 hours and 20 minutes.
This was achieved using the same ABAB dither pattern 
using two sets of 30 exposures and two sets of 20 exposures with
individual times of 120s. Among the 7/8 galaxies detected in [OII]
($\lambda\lambda$3727 {\AA}), this strategy provides a
signal to noise ranging from 4.3 to 19.9 with an average value of 10.6.

The data reduction was performed in two steps. A detailed description
of this reduction can be found in \citet{nanayakkara16}, and we
describe it only briefly here. 

First, a custom version
of the public MOSFIRE data reduction pipeline from the 2015A semester
was used to reduce the raw data into individual 2D spectra for each
object. This step includes flat-fielding, thermal background
subtraction (K band), wavelength calibration, a barycentric correction
to the wavelength solution, careful sky-subtraction following
\citet{kelson03}, and rectification. The calibration provides spectra
with vacuum wavelengths with a residual error of <0.1 {\AA}.

Next, a flux calibration was performed using a custom IDL package. For
our study, an accurate flux calibration is essential for
measuring the emission line ratio of [OIII] to
[OII] given they fall in separate
bands. A  standard star HIP 43018 of A0V type was observed in the
MOSFIRE long-slit ``long2pos'' mask with a slit width of 0$\farcs$7.
We used the standard star for both the telluric correction and flux
calibration. Note that the seeing during the standard star observation was
$\sim$0$\farcs$8 in both the H and the K band, yielding a constant absolute
flux calibration offset due to slit-loss. Slit-loss from the standard star does not affect the
line ratio measurements between the H and K bands because the flux
offset in each band is constant. Hydrogen
absorption lines are removed from the standard star spectra using Gaussian
fits and the resulting spectra are divided by a blackbody function
with the temperature given by the expected temperature of the standard
star. The resulting spectra are normalised, smoothed, and are assumed
as the sensitivity curves of our H and K band observations. We apply these curves to
the standard star observations, then compare the flux density in the given band to
the expected values from the 2MASS catalog \citep{skrutskie06}, thus
calculating the factor used to convert data-units to flux-units. 

Finally, the difference in seeing between our K and H
band observations (0$\farcs$79 vs 0$\farcs$96) may result in a larger
portion of the emission line flux being blurred outside of our
0$\farcs$7 slit for our H band observations (i.e. larger slit
loss for galaxy observations). This results in a relative underestimation of the [OII]
($\lambda\lambda$3727 {\AA}) flux when compared to [OIII]
($\lambda$5007 {\AA}), and thus an overestimation of the [OIII]
($\lambda$5007 {\AA})/[OII] ($\lambda\lambda$3727 {\AA}) ratio. We
make a quantitative estimate of this effect for each galaxy by
degrading the PSF of the HST F814W images to the K and H band values
of 0$\farcs$79 and 0$\farcs$96. We then compare the
fraction of the flux falling within a pseudoslit matching our
MOSFIRE slitmask. We find that, in order to compensate for the
different fractions of the source falling inside the slit, the flux
from the 0$\farcs$96 image (corresponding to H band
observations) must be scaled by a factor of 1.12-1.25 depending on
galaxy morphology and slit position. The slit loss correction values
are quoted in Table \ref{table:2}. We apply this correction to the
measured [OII] ($\lambda\lambda$3727 {\AA}) line fluxes prior to our
correction for dust attenuation.

\subsection{LRIS}

LRIS observations for our MOSFIRE sample come from a larger sample of
22 galaxies selected in a similar method as that
described in Section \ref{section:colsel}. These galaxies were
observed on 19 March, 2015. The goal of our
companion LRIS pilot program is the detection of LyC emission from
star-forming galaxies at $z=3.0$ and higher. 

For our observations the instrumental setup was as follows. Our LRIS
slit mask employed 1$\farcs$2 slits and we use the 560 dichroic
filter. Spectra were dispersed using the 400/3400 and 400/8500
gratings for the blue and red arm of LRIS, respectively. Each mask was
observed with a series of 16 exposures. For the blue arm exposure
times were 1200s giving a total time of 5.33 hrs and for the red arm
exposure times were 1131s giving a total time of 5.03 hrs. The average
seeing for the sample presented here had a FWHM of $\sim$1$\farcs$0.

The LRIS data reduction was performed using the standard IRAF software 
procedures, the basic steps are as follows. First, a conversion is
performed using the multi2simple task in the KECK.LRIS IRAF package,
and during this process the overscan region is removed and bias
corrections are performed. Master science frames for each target are
produced by averaging all single exposures using imcombine and each
master frame is flat fielded using a master flat produced from our
twilight spectra. Next, apall is used to extract the 1D from the 2D
spectra, and background subtraction is also performed during this
step. We then apply a wavelength solution extracted from arc lamp
observations. Finally, we apply a flux calibration based on standard
star observations performed on the same night as our science
observations. 

Analysis of the full LRIS dataset is ongoing. More details on our data
reduction process and a full spectral analysis of LyC emission in our
full sample will be the subject of future work (Me\v{s}tri\'{c} et
al. in preparation). In this work, Ly$\alpha$ properties of our LRIS spectra are used to produce
matched template spectra for estimating $f_{esc}(LyC)$ from
photometric observations in Section \ref{section:fescphot}. A careful
assessment of LyC emission in stacked LRIS 
spectra will be included in Me\v{s}tri\'{c} et al. (in preparation).

\section{Analysis}\label{section:analysis}

\subsection{Optical Emission Line Fitting}\label{section:elf}

\begin{figure}
  \includegraphics[width=\columnwidth]{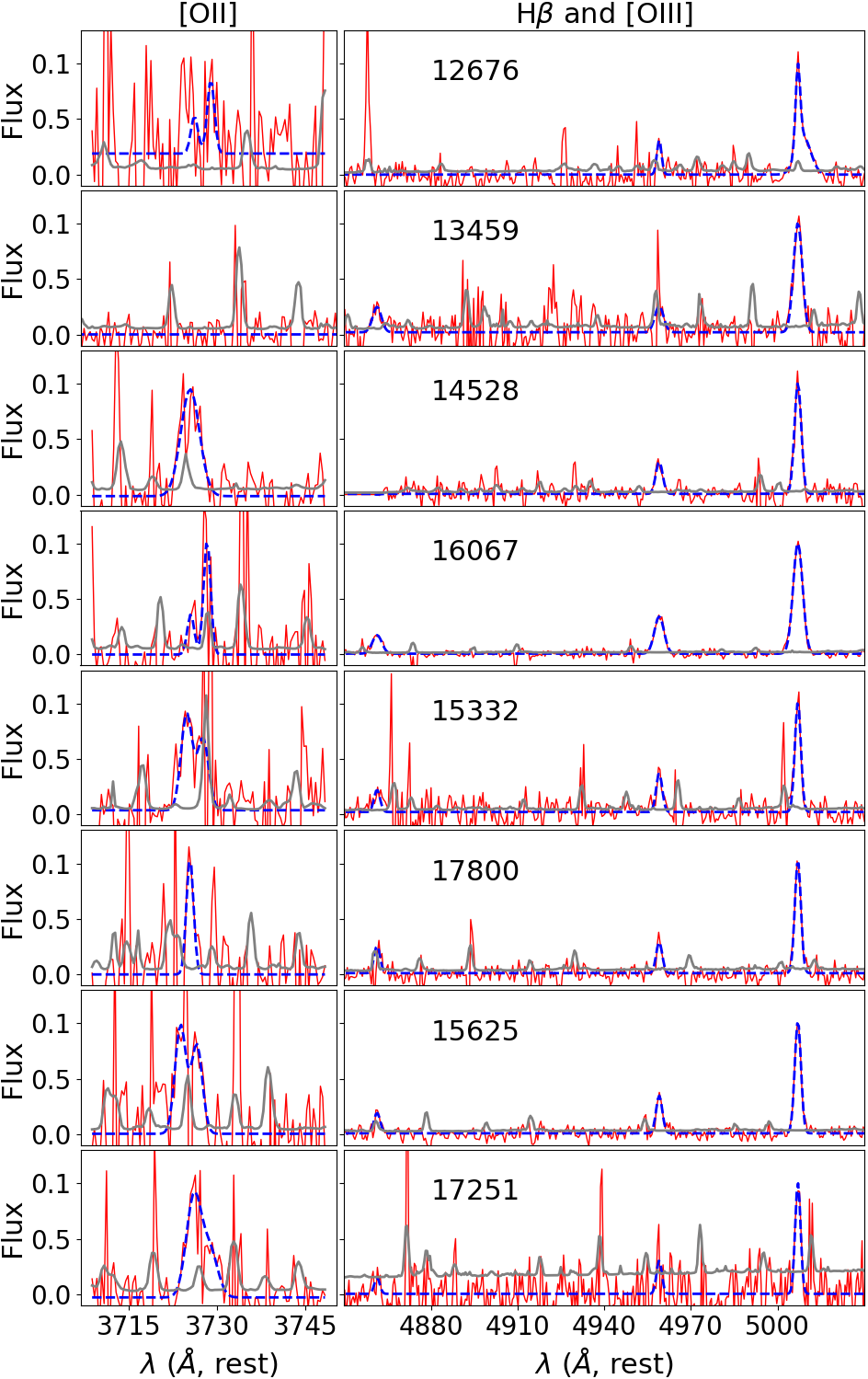}
  \caption{MOSFIRE spectra for $uS$ selected $z\sim3$ galaxies
    in our sample. Red lines show the observed spectra whereas blue
dashed lines show our emission line fits as described in Section
\ref{section:elf}. The grey lines show the error spectrum. For all
galaxies shown here, the [OII]
doublet is observed in the H-band while H$\beta$ and the [OIII]
doublet both fall in the K-band. H-band data are plotted normalised by the
maximum flux in our [OII] line fits, and, similarly, K-band data are
plotted normalised by the peak flux of our [OIII] and H$\beta$ line fit. Dust
corrected line fluxes are given in Table \ref{table:2}.}
  \label{fig:lfs}
\end{figure}

For the MOSFIRE H and K bands we simultaneously fit all emission lines in that band
using Gaussian profiles. The free parameters in each fit are the peak
flux densities at the centres of each line ($A_{\lambda}$), the continuum level, a fixed line
width, and the
redshift, where the latter two values are
assumed to be the same for all lines. Although
none of our galaxies have continuum detected in our spectra, the
continuum level in our fit accounts for any residual zero-point offsets from our data
reduction. Our fits are 
thus performed with five and six free parameters for H and K band 
respectively. Finally, we note that galaxy 12676 exhibits a complex
line shape for [OIII] ($\lambda$5007 {\AA}), possibly due to the
multiple components seen in the HST imaging in Figure
\ref{fig:HST}. Thus, we include a second
Gaussian component for this line allowing the line width and
redshift to vary relative to the other components. This approach has only a
minor effect on the measured line flux and negligible effect on our
resulting emission line ratios. MOSFIRE emission line fits are illustrated in Figure
\ref{fig:lfs}, where reduced 1D spectra are shown in red and our
emission line fits are shown in blue. Line fluxes for each line are calculated
from our Gaussian fits as:
\begin{equation}
  F(\lambda) = \sqrt{2\pi}A_{\lambda}\sigma
\end{equation}

Figure \ref{fig:lfs} illustrates that, in the H band,
the [OII] doublet is not well resolved. In most cases both
Gaussian components are required to provide a reasonable fit to the
data. In cases where the [OII] doublet is well fit by a single
Gaussian profile, we take the flux of this single Gaussian as the
total [OII] ($\lambda \lambda$3727 {\AA}) flux. Thus, regardless of
the relative flux of the two lines, our fitting procedure accounts for
the total flux of the observed doublet.

We calculate the error of our emission line fluxes by combining the
observed spectra with the error spectrum extracted during our data
reduction. For each spectrum we create 1000 realisations of possible
noise spectra by selecting random values in each spectral bin from a
Gaussian distribution with a $\sigma$ equal to the error spectrum in
that bin. We then combine each noise spectrum with the best fit
spectra to create 1000 artificial observations. We then refit the
spectral lines in these artificial observations, and the error in flux
for each line is taken as the standard deviation of computed
values.

We also make a rough estimate of the upper limit of the [OII]
($\lambda\lambda$3727 {\AA}) flux for galaxy 13459 allowing for limits
to be placed on emission line ratios in Section
\ref{section:linerats}. The upper limit here is simply taken as three times the
gaussian noise in the 1D H-band spectrum for this galaxy. We do not
include similar limits on the H$\beta$ flux for galaxies 12676 and
14528 as the former is contaminated by strong skyline residuals and
the latter has H$\beta$ falling below the spectral coverage of our
observations.

From Figure \ref{fig:lfs} galaxy 16067 appears to have a very
large ratio of the [OII] doublet ($\lambda$3729 {\AA}/$\lambda$3726
{\AA}), which has been shown by \citet{sanders16} to have a maximum
theoretical value of $\sim$1.45. For galaxy 16067 we obtain a value of
1.96, significantly higher than the theoretical limit. We randomly generate
10,000 simulated line ratios for this galaxy where each time both line fluxes
are independently varied within a normal
distribution having a $\sigma$ defined by the computed error (as
described above). Among the simulated line ratios, only $\sim$4\% fall
below the theoretical upper limit, thus the computed doublet ratio for
galaxy 16067 is unreasonably large, even within our measured
uncertainties. The most likely cause is significant OH emission
overlapping with the [OII] ($\lambda$3729 {\AA}) line. We retain the
measured, though implausible, value in our analysis with the caveat
that the total [OII] flux is likely an overestimate. The ultimate
result of this overestimate is that the O32 measurement for galaxy
16067 is likely underestimated, a caveat that we carry throughout this
work. Overall, this would not alter the conclusions of this work. We
also note that the [OII] doublet ratios of the remaining galaxies
where both lines are well resolved are
well within the theoretical limits presented in \citet{sanders16}. 

\subsection{Attenuation Correction}\label{section:dust}

\begin{figure}
  \includegraphics[width=\columnwidth]{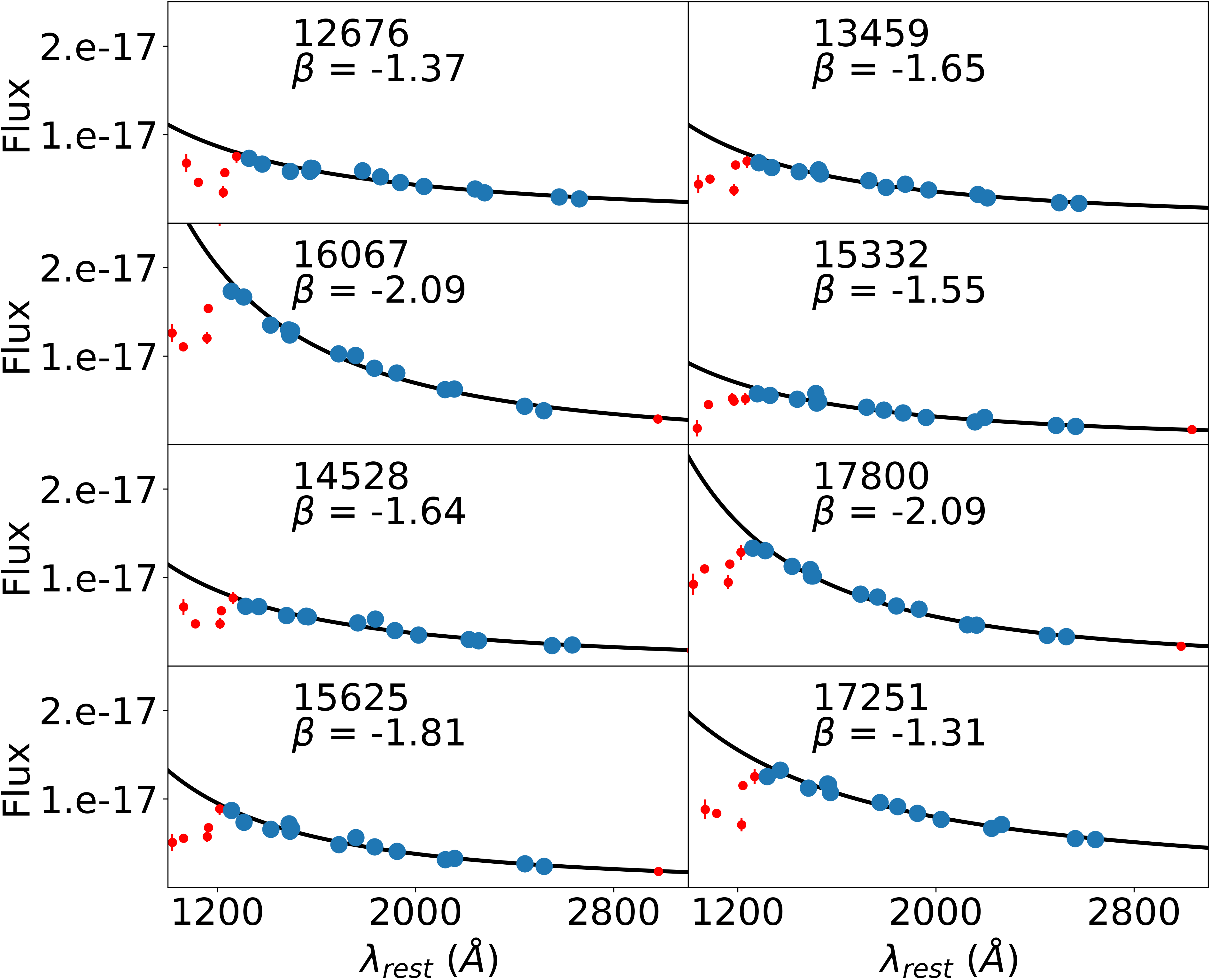}
  \caption{Exponential fits to the restframe photometric data
    (ZFOURGE) for galaxies in our sample. The y-axes indicate
    flux in units of ergs s$^{-1}$ cm$^{-2}$ {\AA}$^{-1}$. Model fits,
    indicated by the solid black 
    lines, are of the form $F(\lambda)\propto
    \lambda^{\beta}$. Data from photometric bands used in our fits are
  shown in blue while data below restframe $\sim$1200 {\AA} (shown in red), below
  which the galaxy SED deviates significantly from an exponential
  behavior as a result of Ly$\alpha$ absorption. The final fitted values of $\beta$ are
  indicated in each panel.}
  \label{fig:beta}
\end{figure}

Line fluxes for each galaxy are corrected for internal dust
attenuation, which is estimated
based on the observed slope of the ultraviolet (UV) continuum
emission, $\beta$. Here, fits are of the form
$F(\lambda)\propto\lambda^{\beta}$. This is a common technique in the literature based on the
IRX-$\beta$ relation of \citet{meurer99}. For our sample, we measure
the slope of the rest-frame UV light from the ZFOURGE photometry in
the wavelength range from $\sim$5200-10500 {\AA} corresponding to
rest-frame wavelengths of $\sim$1250-2500 {\AA} at the redshift of our
sample. This restframe wavelength used for fitting to $\beta$
range matches that recommended by \citet{calzetti94}. For COSMOS data,
this includes 14 photometric bands for 
performing our fit to $\beta$. UV continuum fits for galaxies in our
MOSFIRE sample are shown in Figure \ref{fig:beta}. 

Next, we must convert the observed $\beta$ values to a reddening,
E(B-V), in order to apply a dust correction at longer
wavelengths. Following \citet{reddy17} we assume
the Small Magellanic Cloud (SMC) attenuation curve from \citet{gordon98} and
an intrinsic UV continuum slope of $\beta_{0}=-2.616$. This is a
steeper UV continuum slope than is typical of low redshift 
star-forming galaxies, however there is growing evidence that such a
steep slope may be common at high redshift. \citet{reddy17} show that
together, these assumptions result in a relationship of:
\begin{equation}
  E(B-V) = 0.232+0.089\times\beta
\end{equation}
which we use to calculate E(B-V) for each of our galaxies. 

We also
assume that  the
nebular and stellar attenuation are equivalent rather than applying
the correction of E(B-V)$_{*}$~=~0.44$\times$E(B-V)$_{gas}$ typically
applied to low redshift star-forming galaxies \citep{calz00}. Recent works have shown
that attenuation in highly star-forming galaxies at both high and low
redshift are better described by the relation E(B-V)$_{*}$ =
E(B-V)$_{gas}$ \citep{erb06,reddy10,kashino13,reddy15,bassett17}. 

Finally, the resulting E(B-V) is then applied, along with our chosen SMC
attenuation curve, to the observed optical emission line fluxes. The
final, dust corrected emission line fluxes, as well as the measured
$\beta$ and E(B-V) values, are shown in Table \ref{table:2}. Given the
low levels of attenuation found for galaxies in our sample, assuming
different dust curves \citep[e.g.][]{calz00,card89} or applying a
factor of 0.44 to convert stellar E(B-V) to a nebular value will have
little effect on our final results.

\subsection{Photoionization Modeling}\label{section:pmod}

We calculate a variety of photoionization models using the publicly
available MAPPINGS~V code \citep[see][for the most recent publication
regarding a previous release]{allen08}. MAPPINGS~V is a photoionization
modeling code that performs one dimensional radiative transfer
calculations in order to model the emission line fluxes of various
astrophysical sources including H\,{\sc ii} regions, planetary nebulae, and
active galactic nuclei. MAPPINGS V inputs include a source spectral
energy distribution (SED), nebular abundances, physical parameters
(e.g. density, pressure, ionization parameter, etc.), atomic line data
from the laboratory, and dust depletion data (among others). 

For this work we calculate two types of H\,{\sc ii} region models for comparison with our
MOSFIRE sample. Specifically, these represent ionization bounded and
density bounded H\,{\sc ii} regions. We describe these two classes of models
in the remainder of this Section. The
relevant outputs taken from
MAPPINGS V for our work are the optical emission lines [OII]
($\lambda\lambda$3727 {\AA}), [OIII] ($\lambda$4959 {\AA}), [OIII]
($\lambda$5007 {\AA}),
and H$\beta$, which we compare with the values observed in our
sample. 

Ionization bounded models represent H\,{\sc ii}
regions in which all ionizing photons produced by the central
source(s) (e.g. LyC photons) are absorbed and re-emitted at longer
wavelengths as a variety of optical and IR emission lines. This
type of model results in a shell-like structure, with highly ionized
gas at the centre surrounded by a shell of neutral gas. Ionization
bounded models are very commonly assumed for H\,{\sc ii} regions in low
redshift 
star-forming galaxies and correspond to the conditions often
referred to as Case-B recombination \citep{osterbrock89}. 

\begin{table*}
  \begin{center}
  \caption{MOSFIRE Sample: Optical Line Fluxes}
  \vspace{2mm}
  \begin{tabular}{ c c c c c c c c c }
    \hline\hline
    ID & [OII] ($\lambda\lambda$3727)$^{a,b,c,d}$ & slit corr. & H$\beta^{a,b,c}$ & [OIII] (4959) $^{a,b,c}$ &
                                                                  [OIII]
                                                                  (5007) $^{a,b,c}$
    & $z_{spec}^{e}$ & $\beta^{f}$ & E(B-V) \\
    \hline
12676 & 2.22$\pm$0.18 (0.98) & 1.25 & -- & 1.45$\pm$0.11 (0.94) &
                                                                  4.97$\pm$0.36
                                                                  (3.24)
    & 3.07 & -1.37 & 0.11 \\
13459 & $\leq$0.67 (0.41) & 1.20 & 1.34$\pm$0.11 (0.93) &
                                                          1.36$\pm$0.22
                                                          (0.96) &
                                                                   5.70$\pm$0.18
                                                                   (4.01)
    & 3.09 & -1.65 & 0.09\\
14528 & 11.47$\pm$1.39 (5.71) & 1.24 & -- & 2.92$\pm$0.09 (2.05) &
                                                                   10.03$\pm$0.14
                                                                   (7.07)
    & 3.00 & -1.64 & 0.09 \\ 
15332 & 5.5$\pm$2.1 (3.03) & 1.12 & 1.19$\pm$0.1 (0.83) &
                                                          2.08$\pm$0.11
                                                          (1.46) &
                                                                   5.83$\pm$0.2 (4.1) & 3.11 & -1.55 & 0.09 \\
15625 & 4.78$\pm$0.87 (2.67) & 1.23 & 1.62$\pm$0.25 (1.22) &
                                                             2.98$\pm$0.12
                                                             (2.27) &
                                                                      8.82$\pm$0.15 (6.72) & 3.18 & -1.81 & 0.07 \\ 
16067 & 3.71$\pm$0.37 (2.39)$^{g}$ & 1.19 & 3.95$\pm$0.15 (3.23) &
                                                             7.89$\pm$0.16
                                                             (6.48) &
                                                                      23.37$\pm$0.21 (19.24) & 3.19 & -2.09 & 0.05\\ 
17251 & 14.6$\pm$3.01 (6.5) & 1.18 & 0.42$\pm$0.09 (0.26) &
                                                            0.85$\pm$0.16
                                                            (0.53) &
                                                                     2.82$\pm$0.23 (1.77) & 2.99 & -1.31 & 0.12 \\ 
17800 & 2.32$\pm$0.5 (1.53) & 1.16 & 1.58$\pm$0.25 (1.29) &
                                                            1.79$\pm$0.14
                                                            (1.47) &
                                                                     6.49$\pm$0.2 (5.34) & 3.17 & -2.09 & 0.05 \\ 
    \hline
  \end{tabular}\label{table:2}\\
  \end{center}
  Notes:\\ $^{a}$Emission line fluxes are in units of 10$^{-17}$ ergs
  s$^{-1}$ cm$^{-2}$\\
  $^{b}$Corrected for internal extinction based
  on E(B-V) calculated from the UV slope $\beta$\\ assuming a SMC
  attenuation curve and E(B-V)$_{stars}$ = E(B-V)$_{gas}$.\\
  $^{c}$Non-dust corrected fluxes given in parentheses.\\
  $^{d}$[OII] ($\lambda\lambda$3727) multiplied by value in next
  column to account for slit loss as described in Section \ref{section:mosfobs}\\
  $^{e}$Spectroscopic redshift from fit to K band data. We
  estimate an uncertainty of $\Delta z=0.0011$, the average
  difference between the K and H band spectroscopic redshifts for our sample.\\
  $^{f}$UV continuum slope measured at $\lambda_{rest}\simeq$1250-3000
  {\AA}.\\
  $^{g}$[OII] ($\lambda\lambda$3727 {\AA}) possibly overestimated due
  to OH contamination, see Section \ref{section:elf}.
\end{table*}

We calculate
emission line fluxes for ionization bounded nebulae using MAPPINGS V,
with a fixed, isobaric gas pressure of $P/k=10^{5}$ cm$^{-3}$ K and a
fixed gas temperature of 10$^{4}$ K. This can be converted to a
constant H\,{\sc ii} density, $n_H$, as $n_H\propto P/T$. The
proportionality factor varies slightly with metallicity resulting in
densities of 
$n_H\simeq3.9$ cm$^{-3}$ at low metallicity and $n_H\simeq6.5$ cm$^{-3}$
at high metallicity (corresponding electron densities, $n_{e}$, are
4.0 cm$^{-3}$ and 6.8 cm$^{-3}$). We assume a two-sided,
plane-parallel geometry and a gas filling factor of 1.0. 
Grids of models are calculated by varying the ionization parameter,
$q_{ion}$, defined as:
\begin{equation}
  q_{ion} = \frac{Q_{H^{0}}}{4\pi R_{s}^{2}n_{H}}
\end{equation}
where $Q_{H^{0}}$ is the number of hydrogen ionizing photons produced
per second, $R_{s}$ is the Str\"{o}mgen radius, and $n_{H}$ is the
density of hydrogen. In this work we vary $q_{ion}$
from $log_{10}(q_{ion})$ = 6.5 to $log_{10}(q_{ion})$ = 8.5 and varying metallicity from
$12+log_{10}(O/H)=7.86$ to $12+log_{10}(O/H)=8.99$. The source of ionizing
photons in all models is an SED of an instantaneous star-burst
computed using STARBURST99 \citep{leitherer99} and assuming a \citet{salpeter55} initial
mass function (IMF) with masses of individual stars in the range 0.1-100
M$_{\odot}$. We compute models at metallicities of 0.001 < $Z$ < 0.04,
where stellar and nebular metallicities are matched similar to
\citet{nicholls17}, and SEDs are sampled at a starburst age of 1 Myr. 

Density
bounded nebulae, on the other hand, are models in which the flux of
ionizing photons from the central source is so large that the gas
between the source and the observer is fully ionized. In these types
of models, ionizing radiation is able to escape because there is
little or no
neutral gas remaining between the source and observer to absorb these
photons. Thus, density bounded nebulae models are likely more relevant
to discussion of galaxies with nonzero $f_{esc}(LyC)$. In MAPPINGS V,
we produce density bounded models by specifying an optical depth limit
for hydrogen (HI only), $\tau_{HI}$. In practice this is achieved by
varying the radius at which the output is evaluated at a fixed HI
density. $\tau_{HI}$ is converted to an $f_{esc}(LyC)$ as:
\begin{equation}
  f_{esc}(LyC) = e^{-\tau_{HI}}
\end{equation}
Similar to our ionization bounded models we calculate emission line
fluxes for density bounded models at fixed gas pressure and
temperature. We again vary the ionization parameter from
$log_{10}(q_{ion})$ = 6.5-8.5 with each grid calculated at a fixed
metallicity, varying $\tau_{HI}$ from 200 to 0.001. This effectively corresponds
to a variation in $f_{esc}(LyC)$ from $\sim$0.0 to 0.999 noting
that for $f_{esc}(LyC)$ = 1.0, no emission lines would be present due
to an absence of an ionized nebular region. We calculate three grids at fixed
$12+log_{10}(O/H)$ = 7.86, 8.48, and 8.99. 

Results of our photoionization modeling calculations are presented in Sections
\ref{section:mappings1} and \ref{section:mappings2}. 

\subsection{Estimating $f_{esc}(LyC)$ from Photometry}\label{section:fescphot}

\begin{figure*}
  \includegraphics[width=\textwidth]{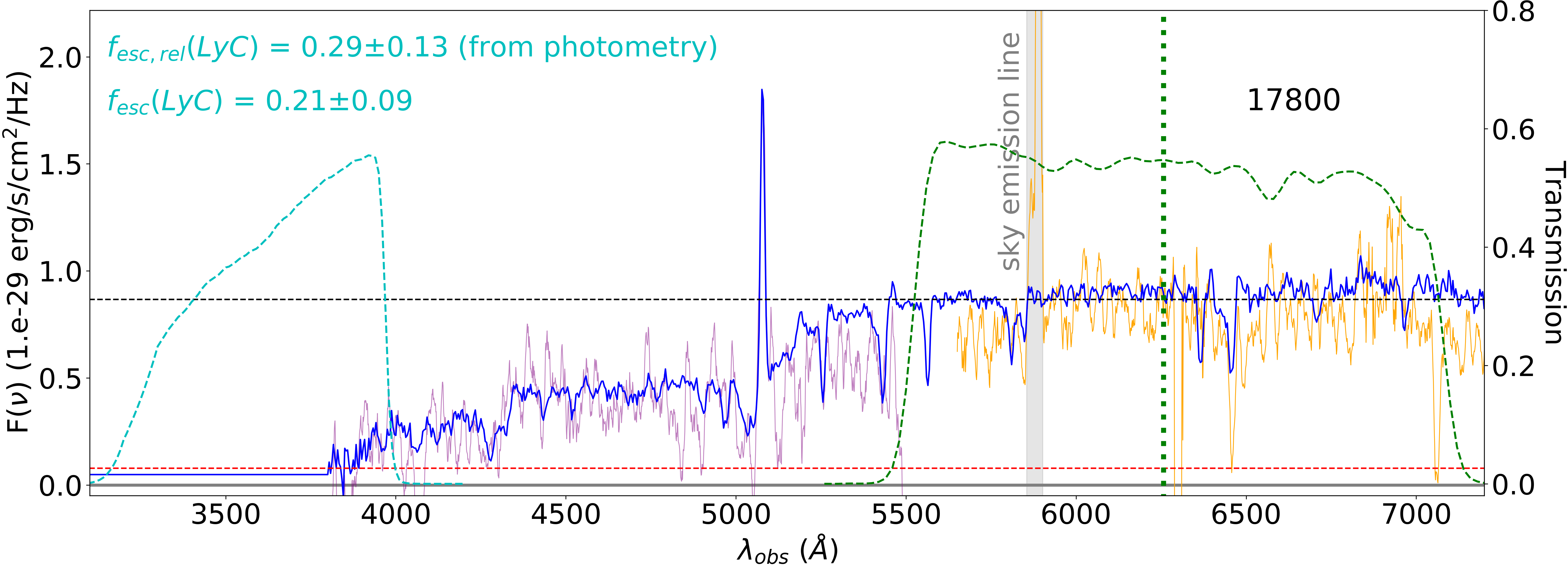}
  \caption{An illustration of our method of estimating $f_{esc}(LyC)$
    from CLAUDS $u$ band photometry for galaxy 17800. CLAUDS $u$ and $r$
    transmission curves (with atmospheric transmission, optics
    throughput, and CCD quantum efficiency accounted for) are shown by cyan and green dashed lines,
    respectively, with transmission given by the right
    y-axis. Corresponding observed $F_{\nu}$ in the $u$ and $r$ bands
    are shown by the horizontal red and black dashed lines,
    respectively. The composite spectrum, constructed from stacked LBG
    spectra of \citet{shapley03}, is normalised based on the $r$ band
    flux and is shown by the solid blue line. Our LRIS spectrum, which has
    been smoothed using a tophat filter with a width of 20 {\AA}, for
    the blue and red arms are shown in purple and orange,
    respectively (note that the quoted $f_{esc}(LyC)$ is \textit{not}
    measured from the LRIS spectrum). The estimated value for photometric $f_{esc}(LyC)$
    is annotated in the upper left. Finally, we indicate $\lambda_{rest}=1500$
  {\AA} with a green, vertical dotted line.}
  \label{fig:fescphot}
\end{figure*}

In this Section we describe our method of estimating $f_{esc}(LyC)$
from our photometric measurements. Our process relies on a comparison
of CLAUDS $u$ band and $r$ band fluxes that are extracted using
identical apertures. The contribution to the
observed $u$ band flux from $\lambda$ $>$ 912 {\AA} photons is
modelled using linear combinations of stacked LBG spectra from
\citet{shapley03}. This process, which is necessitated by the fact
that Ly$\alpha$ forest light also contributes to the observe $u$ band
flux at the redshift of our sample, is illustrated in Figure
\ref{fig:fescphot}. 

The first step of this process is to construct a composite spectrum
from stacked LBG spectra of \citet{shapley03} that best matches
the Ly$\alpha$ emission in our observed LRIS spectrum, noting that 5/8
galaxies in our sample exhibit Ly$\alpha$ in emission. 
Comparing stacked LBG spectra with and without Ly$\alpha$ emission,
LAEs are known to have a bluer UV
continua \citep{cooke14} and thus, similarly higher Ly$\alpha$ forest flux relative
to the UV continuum normalised at 1500 {\AA} when compared to LBGs
with Ly$\alpha$ in absorption. These properties vary systematically with
Ly$\alpha$ EW. Thus, a mismatch between our observed
LRIS spectra and the Ly$\alpha$ properties in composite spectra can
affect our estimates of $f_{esc}(LyC)$ from photometric
observations. For our analysis we use a total of four composite
spectra: one Ly$\alpha$ absorber and three composite spectra with increasing
Ly$\alpha$ EW where composite spectra are constructed as linear
combinations of stacked LBG spectra from \citet{shapley03} that best
match the Ly$\alpha$ EW observed in our LRIS spectra. We test for the
level of variation in $f_{esc}(LyC)$ that may be 
attributed to such a mismatch by estimating $f_{esc}(LyC)$ using
all four composite spectra. For most galaxies we find a variation of
$\Delta f_{esc}(LyC)$ $<$ 0.05
among all four composite spectra. Our lowest redshift galaxy, galaxy 17251, has a larger
variation of 0.09 due to a larger contribution to the $u$ band from
the Ly$\alpha$ forest.

Next, the composite spectrum for each galaxy is normalised based on the
observed $r$ band flux in $F_{\nu}$. At the average redshift of our
sample of $\langle z \rangle =3.17$, the $r$ band samples the rest-frame UV continuum
around 1500 {\AA}, which is the wavelength typically considered for
measuring LyC escape \citep[e.g.][]{steidel18}. The normilisation factor for our
composite spectrum is taken as the observed $r$ band $F_{\nu}$, $F_{\nu}(UV)$, divided
by the median $F_{\nu}(UV)$ of the matched composite spectrum for a given
galaxy. The matched composite spectrum is then multiplied by the
normalisation factor. In Figure \ref{fig:fescphot} 
we show the observed $F_{\nu}(UV)$ with the black dashed line and
the normalised composite spectrum with the solid blue line. The
resulting spectral regions blueward of 
Ly$\alpha$ is assumed to match the underlying spectra of our galaxy
sample. This appears reasonable as we find a good agreement between
normalised composite spectra and the observed $u$ and $g$
magnitudes. For comparison, we also show an observed LRIS spectrum,
which 
has been smoothed using a tophat function with a width of 20 {\AA},
and it can be seen that the composite spectrum is indeed well matched to
the observed spectrum at blueward of Ly$\alpha$. Finally, we note that
at observed wavelengths below restframe 912 {\AA}, composite spectra
have a flux density of 0.

Having produced a normalised composite spectrum for each galaxy, we
then estimate the expected $F_{\nu}(u-band)$, the modeled flux density in
the $u$ band, assuming zero flux below 912 {\AA} (i.e. $f_{esc}(LyC)=0.0$). This value is
taken as the weighted average of the composite spectrum, redshifted
based on the observed redshift from our MOSFIRE emission lines, where the
weights are given by the $u$ band sensitivity. This weighting takes
into account the CLAUDS $u$ transmission curve, the detector quantum
efficiency (QE), optical throughput of the telescope/instrument, and
atmospheric transmission. The resulting transmission curve is shown as
a cyan dashed line in Figure \ref{fig:fescphot}, and shows that
including these additional effects (QE in particular) produces a
strong weighting towards the redder portion of the filter (compare to
the filter only transmission in Figure \ref{fig:lcgspec}). As the red
portion of the filter contains the Ly$\alpha$ forest contribution,
accounting for the full system response is critical in accurately
estimating $f_{esc}(LyC)$.

After computing the expected $F_{\nu}(u-band)$
assuming $f_{esc}(LyC)=0.0$, we then compare this to the observed
value from CLAUDS photometry indicated by the red, dashed line in
Figure \ref{fig:fescphot}. For three galaxies the modeled
$F_{\nu}(u-band)$ for $f_{esc}(LyC)=0.0$ is lower than the observed value. 
This indicates that additional flux, beyond what is contained in the
Ly$\alpha$ forest, is required in the $u$ band to match the
photometric observations (i.e. $F_{\nu}(LyC)$). We show
the observed $F_{\nu}(u-band)$ in Figure \ref{fig:fescphot} using the red
dashed line. We note here that, due to the low sensitivity to LyC
photons in the $u$-band observations (indicated by the sensitivity
curve in Figure \ref{fig:fescphot}), the modeled values here may be
somewhat underestimated.

For cases where additional flux is required in our composite
spectrum to match the observed $F_{\nu}(u-band)$, we then iteratively add
flux to the composite spectrum at $\lambda_{rest}<$ 912 {\AA} until the
modeled $F_{\nu}(u-band)$ matches the observed value. This added flux in
the LyC region is added simply as a flat distribution in
$F_{\nu}(\lambda)$ over $u$-band wavelengths where the amount of flux added in each iteration is
equal to 0.01 times the observed $F_{\nu}(UV)$. Thus, our estimate
of $F_{\nu}(LyC)$/$F_{\nu}(UV)$ has a precision of 0.01, which we find
equates to an $f_{esc}(LyC)$ 
precision significantly smaller than the corresponding error due to
uncertainty in the photometric measurement.

In cases where no additional flux is required to match the observed
$F_{\nu}(u-band)$, we then estimate the upper limit on $f_{esc}(LyC)$ based
on the photometric uncertainty of the CLAUDS $u$ band
observation. This is done by adding the uncertainty in $u$ to the
observed value and again comparing to the modeled $u$ flux for
$f_{esc}(LyC)=0.0$. In cases where we still find no additional flux is required,
we record an $f_{esc}(LyC)=0.0$ for that galaxy. Otherwise, we perform
the same iterative process described above, this time using the
observed $F_{\nu}(u-band)$ with the added photometric uncertainty. In such
cases, the resulting $f_{esc}(LyC)$ value is taken as an upper
limit. The blue composite spectrum 
in Figure \ref{fig:fescphot} illustrates the flat $F_{\nu}(LyC)$
required for galaxy 17800 such that the modeled $F_{\nu}(u-band)$ matches
the observed value with the photometric uncertainty added. 

Next, we must convert the estimated
$F_{\nu}(LyC)$ to $f_{esc}(LyC)$ for each galaxy. To do this 
requires an assumption regarding the intrinsic
LyC to UV luminosity, $(L_{\nu}(LyC)$/$L_{\nu}(UV))_{int}$, and the transmission of LyC through the
IGM at the redshift of each galaxy. While there may be some variation
in $(L_{\nu}(LyC)$/$L_{\nu}(UV))_{int}$, particularly at high redshift
\citep[e.g.][]{chevallard17}, we assume a value of 0.33, which is a commonly
assumed value in the literature
\citep{steidel01,inoue05,shapley06,vanzella12}. We check how
reasonable this value of $(L_{\nu}(LyC)$/$L_{\nu}(UV))_{int}$ is for
our sample by estimating the efficiency of ionizing photon production,
$\xi$ as:
\begin{equation}
  \xi = \frac{N_{LyC}}{L(UV)}
\end{equation}
where $N_{LyC}$ is the production rate of ionizing photons. We
calculate $N_{LyC}$ based on the dust corrected H$\beta$ luminosity and the
value of $f_{esc}(LyC)$ (calculated below):
\begin{equation}
  N_{LyC} = 2.1 \times 10^{12} L(\textrm{H}\beta) \times (1 - f_{esc}(LyC))^{-1}
\end{equation}
 This follows \citet{izotov17} using the values quoted in
 \citet{storey95}. We note that estimating $\xi$ also requires 
 a slit correction to $L(UV)$ as this photometric value is taken
 in a circular aperture while H$\beta$ measurement is performed in a
 narrow slit. We can roughly estimate this correction by comparing the
 catalog value of $F(UV)$ to the value measured in our MOSFIRE slit,
 however the true correction may vary if the H$\beta$ emitting region
 is not spatially coincident with the UV emitting region (e.g. if
 H$\beta$ is more extended, this correction will underestimate
 $\xi$). We also note that our attenuation correction for H$\beta$ is
 based on the UV spectral slope of the stellar continuum light
 assuming $E(B-V)_{stars} = E(B-V)_{gas}$. If the nebular emission is
 in fact more attenuated than the stellar light \citep[as is often the
 case at low $z$, e.g.][]{calzetti94,calz00}, our attenuation
 corrected value of L(H$\beta$), and thus $\xi$, will be
 underestimated. 

For H$\beta$ detected galaxies in our sample, values of
 log$(\xi)$ vary from 24.44 to 25.07. These values are 
 lower than the canonical value used for theoretical work on the EOR of
 log$(\xi)\sim25.2-25.3$ \citep[e.g.][]{robertson13}, but are
 comparable to low H$\beta$(EW) compact star-forming galaxies at low
 redshift from \citet{izotov17}. Estimating H$\beta$ EW for our sample
 for comparison is not possible though as our MOSFIRE spectra are not continuum
 detected. Compared to samples in the redshift
 range 1.4 $<$ $z$ $<$ 2.2, our measurements of $\xi$ are lower than
 average, but overlap with the low $\xi$ end of observed galaxies
 \citep[e.g.][]{matthee17,shivaei18}. \citet{nakajima16} estimate
 $\xi$ for LAEs at 3.1 $<$ $z$ $<$ 3.7, a similar redshift to our
 sample, finding log$(\xi)$ $>$ 25 for H$\beta$ detected
 galaxies. Upper limites for H$\beta$ non-detected galaxies from
 \citet{nakajima16}, however, are consistent with the values computed
 for our sample. At $z > 4$, typical $\xi$ measurements are
 significantly higher than our sample
 \citep{bouwens16,stark15,stark17}.

 What may be the cause of our apparent bias of our sample towards low $\xi$
 galaxies? A complex/bursty star-formation
 history and/or a significant old stellar population in our sample
 could explain such low values of $\xi$. Such a complex star-formation
 history may be reasonable given the relatively large stellar masses
 of our sample. Related to this, it has been suggested that $\xi$ may
 be inversely correlated with stellar mass with more massive galaxies
 having a lower production of ionising photons
 \citep[e.g.][]{faisst16,matthee17}, consistent with the low values
 found for our sample. Finally, we note that the
 value of $\xi$ calculated here depends directly on $f_{esc}(LyC)$,
 which itself depends on the assumed $\xi$. Regardless, we adopt a value
 of $(L_{\nu}(LyC)/L_{\nu}(UV))_{int}$ of 0.33 to compare directly
 with other studies. A lower value of $\xi$ corresponds to a lower
 value of $(L_{\nu}(LyC)/L_{\nu}(UV))_{int}$ and, in turn, this would
 result in a \textit{higher} estimate of $f_{esc}(LyC)$. Thus, values
 quoted here should be considered conservative in this sense.

As for
the IGM transmission of LyC, $T_{IGM}(LyC)$, for each galaxy we reproduce the average IGM
transmission curves at the redshift of the galaxy following the
analytic functions provided in \citet{inoue14} and apply the modeled
value at 912 {\AA}.  We then compute the \textit{relative} LyC escape,
$f_{esc,rel}(LyC)$, from our photometric measurements as:
\begin{equation}
f_{esc,rel}(LyC) = \frac{F_{\nu}(LyC)}{F_{\nu}(UV)}\times\left(\left(\frac{L_{\nu}(LyC)}{L_{\nu}(UV)}\right)_{int}T_{IGM}(LyC)\right)^{-1}
\end{equation}
We must then apply a dust correction to the UV luminosity to compute
the absolute $f_{esc}(LyC)$ following \citep{inoue05,siana07}:
\begin{equation}
  f_{esc}(LyC) = f_{esc,rel}(LyC)\times10^{-0.4A_{UV}}
\end{equation}
where $A_{UV}$ is the dust attenuation at 1500 {\AA} evaluated for an
SMC attenuation curve \citep{gordon98} with E(B-V) from the UV slope
as measured in Section \ref{section:dust} (see Table \ref{table:2}).

Due to the stochastic
nature of the column density of neutral hydrogen in any one sightline
through the IGM, the adopted value $T_{IGM}(LyC)$ represents the largest uncertainty
for individual estimates of $f_{esc}(LyC)$. Modeled values of
$f_{esc,rel}(LyC)$ and $f_{esc}(LyC)$ for each galaxy are quoted in Table
\ref{table:fescphot} and we compare these values with O32 ratios in
Section \ref{section:mappings2}. Upper and lower limits on $f_{esc}(LyC)$ are taken by
evaluating $f_{esc}(LyC)$ using the above methodology with the
photometric uncertainty added or subtracted from the observed
value. We note that in most cases subtracting the photometric
uncertainty results in the $u$ mag consisted with
$f_{esc}(LyC)=0.0$. Summarising our assessment of LyC escape in our
sample, 3/8 galaxies have $u$ band flux consistent with 0.07 $<$ $f_{esc}(LyC)$
$<$ 0.37, 2/8 have upper limits on $f_{esc}(LyC)$ of 0.04 and 0.06,
and the remaining 3/8 have $u$ band photometry consistent with
$f_{esc}(LyC)=0.0$.

Although we have obtained LRIS spectra for each galaxy, we defer
estimating $f_{esc}(LyC)$ from these spectra to future work (Me\v{s}tri\'{c} et
al. in preparation) due to the low signal to noise (S/N) of these
spectra. In the LyC region of our LRIS spectra we obtain an average
S/N of $\sim$0.2 per pixel and LyC emission is also often
affected by individual noise spikes in our spectra as seen in Figure \ref{fig:fescphot}. Future work,
relying on an advanced flat-fielding technique and stacking analysis of our larger sample of LRIS observed
galaxies, will provide spectroscopic measurements of the average
$f_{esc}(LyC)$ of $u$ band selected targets (Me\v{s}tri\'{c} et al. in preparation). 

\begin{table}
  \begin{center}
  \caption{Estimates of $f_{esc}(LyC)$ and Ly$\alpha$ EW}
  \vspace{2mm}
  \begin{tabular}{ c c c c }
    \hline\hline
    ID & $f_{esc,rel}(LyC)^{a}$ & $f_{esc}(LyC)^{b}$ & Ly$\alpha$ EW$^{c}$ \\
    \hline
    12676 & $\leq$0.09 & $\leq$0.04 & -- \\ [1ex]
    13459 & 0.0$\pm$0.0 & 0.0$\pm$0.0 & -- \\ [1ex]
    14528 & 0.0$\pm$0.0 & 0.0$\pm$0.0 & 104$\pm$35\\ [1ex]
    15332 & 0.12$^{+0.26}_{-0.12}$ & 0.07$^{+0.14}_{0.07}$ & 50$\pm$11\\ [1ex]
    15625 & 0.0$\pm$0.0 & 0.0$\pm$0.0 & 20$\pm$5\\ [1ex]
    16067 & $\leq$0.09 & $\leq$0.06 & 58$\pm$11\\ [1ex]
    17251 & 0.81$\pm0.16$ & 0.37$\pm0.08$ & 26$\pm$8\\ [1ex]
    17800 & 0.29$\pm0.13$ & 0.21$\pm0.09$ & --\\ [1ex]
    \hline
  \end{tabular}\label{table:fescphot}\\
  \end{center}
  Notes:\\ $^{a}$estimated from CLAUDS $u$ band photometry \\
  $^{b}$absolute $f_{esc}(LyC)$ after applying dust correction (SMC)
  to $f_{esc,rel}(LyC)$\\
  $^{c}$Rest-frame Ly$\alpha$ equivalent width for Ly$\alpha$ emitters only.
\end{table}

\section{Results}\label{section:results}

\subsection{Rest-Frame Optical Emission Line Ratios}\label{section:linerats}

\begin{figure}
	\includegraphics[width=\columnwidth]{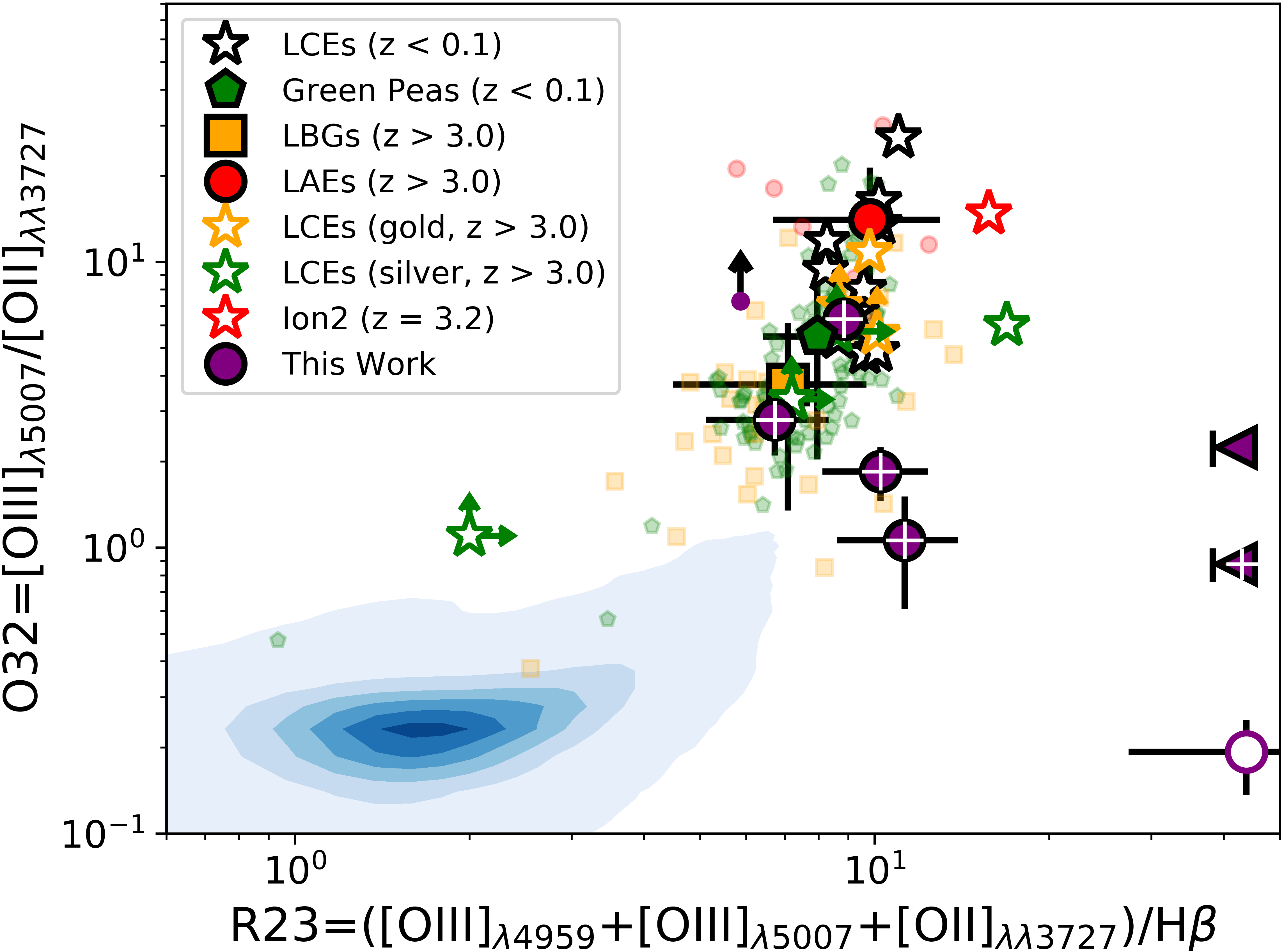}
    \caption{R23 vs O32 for our sample in comparison with various
      samples of high and low redshift star-forming galaxies. For most
      comparison samples, 
      small symbols indicate individual galaxies and corresponding
      large, opaque symbols show averages for each sample. The
      exception is LyC emitting galaxies shown as white stars. Those
      with black edges are low-$z$ galaxies from \citet{izotov16}, \citet{izotov18b}, and
      \citet{izotov18}, the red edged white star is the high-$z$
      galaxy Ion2 \citep{debarros16,vanzella16}, and orange and green edged white
      stars are $z>3.0$ galaxies from the ``gold subsample'' and
      ``silver subsample'', respectively, of the LACES
      survey \citep{fletcher18}. Star-forming SDSS galaxies from
      \citep{zahid13} are shown by the blue density
      contours. Galaxies in our MOSFIRE
      sample with detections of [OII] ($\lambda\lambda$3727 {\AA}),
      [OIII] ($\lambda$4959 {\AA}), [OIII] ($\lambda$5007 {\AA}), and H$\beta$ are
      shown with large purple circles. Circles with white crosses indicate
      confirmed Ly$\alpha$ emitters. Filled triangles represent  those
      galaxies without H$\beta$ observations indicating only O32 values and not R23
      upper-limits. The small purple circle indicates the upper
      limit on O32 for galaxy 13459 (not detected in [OII]
      $\lambda\lambda$3727 {\AA}). The open purple circle represents galaxy
      17251, which shows evidence of being an ongoing merger (see
      Figure \ref{fig:HST} and Section \ref{section:17251}). Shocks
      associated with mergers may explain both the high R23 and low
      O32 values observed in this galaxy.}
    \label{fig:R23O32_1}
\end{figure}

In this Section we present optical emission line ratios for galaxies
with clean detections of [OII] ($\lambda\lambda$3727 {\AA}), [OIII]
($\lambda$4959 {\AA}), and [OIII] ($\lambda$5007 {\AA}) in comparison with
samples of highly ionized galaxies at both low and high redshift. In
particular we examine the line ratios:
\begin{equation}
  R23 = \frac{[OII]_{\lambda\lambda3727}+[OIII]_{\lambda
      4959}+[OIII]_{\lambda 5007}}{H\beta}
\end{equation}
and
\begin{equation}
  O32 = \frac{[OIII]_{\lambda 5007}}{[OII]_{\lambda\lambda3727}}
\end{equation}
R23 has long been
suggested as an indicator of the physical conditions of H\,{\sc ii} regions in
star-forming galaxies \citep{pagel79}. More recently a comparison
between R23 and O32 has been used in studies
of high ionization in star-forming galaxies
\citep[e.g.][]{nakajima14} as O32 is sensitive to the ionization state
of star-forming gas \citep{kewley02,steidel14,steidel16}. Our results are presented in Figure
\ref{fig:R23O32_1}. 

As
discussed in Section \ref{section:elf}, among our selected sample of
eight galaxies at $\langle z \rangle = 3.17$, five were well detected in [OII]
($\lambda\lambda$3727 {\AA}), [OIII] ($\lambda$4959 {\AA}), [OIII]
($\lambda$5007 {\AA}),
and H$\beta$. Of the remaining three, two were only detected in [OII]
($\lambda\lambda$3727 {\AA}), [OIII] ($\lambda$4959 {\AA}), and
[OIII] ($\lambda$5007 {\AA}) and the third was detected in [OIII] ($\lambda$4959 {\AA}),
[OIII] ($\lambda$5007 {\AA}), and H$\beta$. For this latter object
(galaxy 13459) we include an upper limit on [OII] ($\lambda\lambda$3727
{\AA}), but do not include limits on H$\beta$ for the former two
(12676 and 14528) for reasons described in Section \ref{section:elf}
(and see Figure \ref{fig:lfs}). It has been
suggested by recent works \citep[e.g.][]{izotov18} that only
O32 may be required to provide an estimate of $f_{esc}(LyC)$ (see also
Sections \ref{section:mappings1} and \ref{section:mappings2}), thus
we can make an additional estimate of $f_{esc}(LyC)$ independent of
$u$ band photometry even for those galaxies with no H$\beta$
detection. 

The resulting R23 vs O32 values for five galaxies with 
detections in all lines are shown as purple
circles in Figure \ref{fig:R23O32_1}. The five galaxies in our sample that are also found to
be Ly$\alpha$ emitters from our LRIS observations are indicated with 
white crosses. Four out of five galaxies well detected in all lines are found to have a large
average value of 
O32 $\gtrsim$ 1.0, consistent with a high ionization parameter. Ly$\alpha$
emitters are found to cover the full range in O32 ratios observed in
our sample, though the strongest
Ly$\alpha$ emitter in our sample also has the highest O32 of
6.3$\pm$0.7 (excluding the lower limit for 13459). The fifth galaxy from our
sample observed in all five lines (17251), shown with an open purple circle,
exhibits an R23 value inconsistent with pure photoionization as we
describe in Section \ref{section:17251}. 

Galaxy 13459, which is not
detected in [OII] ($\lambda\lambda$3727 {\AA}), is shown as a small
purple circle with an upward arrow indicating this as a lower limit on
O32. This galaxy also has a high O32, indicating high
ionization similar to other galaxies in our sample. Such a large O32
also means that the effect of [OII] ($\lambda\lambda$3727 {\AA}) 
emission on the observed value of R23 will be negligible.

The two galaxies from our observations without a clear detection of
H$\beta$ are plotted as filled purple triangles in Figure
\ref{fig:R23O32_1}. H$\beta$ for these observations either fell below
the wavelength coverage of our observations or overlapped with a
strong skyline (see Figure \ref{fig:lfs}), preventing our measurement.
Thus, the solid purple triangles do not represent
a limit on R23, as this is entirely unknown, but simply serve to indicate their O32
ratio. These two galaxies are found at large O32 with values of
0.9$\pm$0.1 and 2.2$\pm$0.3, in the same range as
for previously discussed galaxies.

Figure \ref{fig:R23O32_1} also displays comparison samples of
low redshift, star-forming galaxies. Blue density contours represent
51,262 star-forming galaxies from the Sloan Digital Sky Survey (SDSS) 
from \citet{zahid13}, and are representative of the bulk of low redshift, star-forming
galaxies. So called ``green pea'' galaxies \citep{cardamone09} are a well-studied population of 
compact and highly star-forming low redshift galaxies. Those presented
in Figure \ref{fig:R23O32_1} are a compilation from \citet{yang16} and 
\citet{cardamone09} galaxies that overlap with the \citet{zahid13}
sample. Compared to the bulk of galaxies from \citet{zahid13}, green
peas are highly ionized, with average O32 values larger than 1.0,
comparable to our sample.

In addition to the low redshift comparison samples, we also show in
Figure \ref{fig:R23O32_1} two redshift $\sim$3-4 samples, typical of
star-forming galaxies at this epoch. These are LBGs compiled from
\citet{troncoso14} and \citet{maiolino08} and LAEs taken from
\citet{nakajima14} and \citet{nakajima16}. LBGs
closely overlap the distribution of green peas whereas LAEs from the literature
are typically found at the highest O32 among all samples in Figure
\ref{fig:R23O32_1}. 

Our MOSFIRE sample is well matched to
the upper end of the O32 distribution for LBGs while falling below
the O32 values of LAEs. This is not surprising as the $(g-r)$ vs
$(u-g)$ colours of our sample are consistent with the standard LBG
selection \citep[e.g.][]{steidel96}. It is interesting to note that,
although LAEs from the literature are found at the very high end of
the O32 distribution, the five galaxies in our MOSFIRE sample that
are also found to emit Ly$\alpha$ (black crosses)
extend this sample to lower O32. This is perhaps not surprising given that
\citet{shapley03} find that roughly half of the LBGs in their
sample exhibit some Ly$\alpha$ in emission. Unlike \citet{shapley03}, 62.5\%
of our selection exhibit Ly$\alpha$ emission, and these emitters have
Ly$\alpha$ EW > 20 {\AA}. Ly$\alpha$ emission of this strength only
occurs in $\sim$25\% of the \citet{shapley03} sample. We also consider
the 13/15 LCE's presented in \citet{steidel18} exhibit Ly$\alpha$ in
emission, however only 4/15 (26.7\%) have Ly$\alpha$ EW > 20
{\AA}. This may suggest that
selecting $z\sim3$ LBGs with high $uS$ fluxes may preferentially
select those with high Ly$\alpha$ EW, though not exclusively. These results highlight the fact that
LAEs and LBGs are likely drawn from the same general parent
sample of high redshift, star-forming galaxies, while their observed
properties simply reflect the differing methods of selection.

A
key comparison for our sample in terms of $f_{esc}(LyC)$, are confirmed
LCEs. Low redshift LCEs taken from
\citet{izotov16}, \citet{izotov18}, and \citet{izotov18b} are
shown as open black stars. These galaxies are selected from a parent sample
of green peas based both on their compactness and their large O32
values. All nine of these galaxies are found to have O32 > 4.0,
and in particular the three strongest LCEs with
$f_{esc}(LyC)=0.38-0.73$, are among the galaxies with the highest
O32 values in the range 11.5-16.3. The remaining low redshift LCEs are found to
have $f_{esc}(LyC)\leq 0.14$. 

Also shown in Figure \ref{fig:R23O32_1} as open red, orange, and green bordered
stars are known samples of $z>3.0$ LCEs also having 
measurements of O32. The red bordered star is the galaxy ``Ion2''
\citep{debarros16,vanzella16} and the orange bordered stars are ``gold
subsample'' galaxies from the LACES survey having measurements of R23 and
O32 \citep{fletcher18}. LACES ``silver subsample'' galaxies,
which have lower S/N detections of LyC, are shown with green bordered
stars, however 3/4 have only lower limits on both R23 and O32. We also
exclude one LACES silver galaxy (galaxy 94460) as they note their spectroscopic
observations indicate possible low redshift contamination. The
LACES gold and silver subsamples include 6 and 4 additional
LCEs, respectively, without optical line ratios.  It should also be
mentioned that all LACES LCEs are found to also be Ly$\alpha$ emitters
with Ly$\alpha$ EW $>$ 20, similar to Ly$\alpha$ emitters in our
sample. All high redshift
LCEs from the literature in Figure \ref{fig:R23O32_1}, excluding lower
limits, have O32 $>$ 5.6, with Ion2 found at the highest O32 of 14.7. 

Taken together, the combined sample of high and low redshift
LCEs with O32 measurements suggest that high ionization, as indicated by a
high O32, is a possible identifying feature of LyC escape from
star-forming galaxies. It should be noted, however, that the
galaxy with the largest O32 value from \citet{izotov18b} of 27.2 has
a relatively low estimated $f_{esc}(LyC)$ value of 0.11. We discuss
the possible connection between observed values of O32 and
$f_{esc}(LyC)$ further in Section \ref{section:mappings2}. 

Among our six galaxies with reliable O32 measurements, one has O32
matching the O32 range exhibited by LCEs. As we have
described, the estimated $f_{esc}(LyC)$ for these low O32 LCEs is on
the order or 0.05-0.17. High redshift LCEs also seem to follow similar trends between
O32 and $f_{esc}(LyC)$ seen by \citet{izotov18}. As we will see in
Section \ref{section:mappings2}, the empirical relationship
between O32 and $f_{esc}(LyC)$ from \citet{izotov18} and
\citet{faisst16} predicts an
$f_{esc}(LyC)$ of 0.10-0.16 for galaxy 16067, our galaxy having the
largest O32 value among those detected in [OII] ($\lambda\lambda$3727 {\AA}). This $f_{esc}(LyC)$ for galaxy 16067 is in
roughly 2$\times$ larger than the upper limit on $f_{esc}(LyC)$ estimated from CLAUDS
$u$ band photometry quoted in Table \ref{table:fescphot}. Furthermore,
as mentioned in Section \ref{section:elf}, the [OII] ($\lambda$3729
{\AA}) flux for galaxy 16067 is contaminated by OH emission and likely
overestimated as indicated by the implausibly large [OII]
doublet ratio ($\lambda$3729 {\AA}/$\lambda$3726 {\AA}). The suggests
an underestimate of O32 that, if true, would imply a larger
discrepancy between the predicted $f_{esc}(LyC)$ (from O32) and the
observed value (from $u$-band photometry). In addition,
we estimate
a lower limit for O32 of 8.51 for galaxy 13459 corresponding to
$f_{esc}(LyC) > 0.22$ based on the relation of
\citet{izotov18}. Interestingly, this galaxy is a non-detection in our
CLAUDS $u$ band image (though a bright, likely low redshift,
$u$ band source can be seen in the cutout shown in Figure
\ref{fig:HST}), and thus consistent with $f_{esc}(LyC) = 0.0$.

From O32 alone, the
remaining galaxies are expected to be negligible emitters of LyC
photons (though we estimate $f_{esc}(LyC)$ from $u$-band
photometry as high as 0.07-0.37 for some of these galaxies, see Sections
\ref{section:fescphot} and \ref{section:mappings2}). We note 
that galaxies in this pilot survey were selected as bright, and thus
likely more massive and less compact compared to Ion2 and green peas,
which may bias us to a sample with lower O32 (and
thus lower $f_{esc}$). See section \ref{section:msfr} for more
discussion of this. A final caveat is that, although all
confirmed LCEs are found at high O32, the value of O32 for other
samples of high redshift LCEs are currently unknown
\citep[e.g.][]{steidel18}. 
  
\subsubsection{Galaxy 17251}\label{section:17251}

\begin{figure}
	\includegraphics[width=\columnwidth]{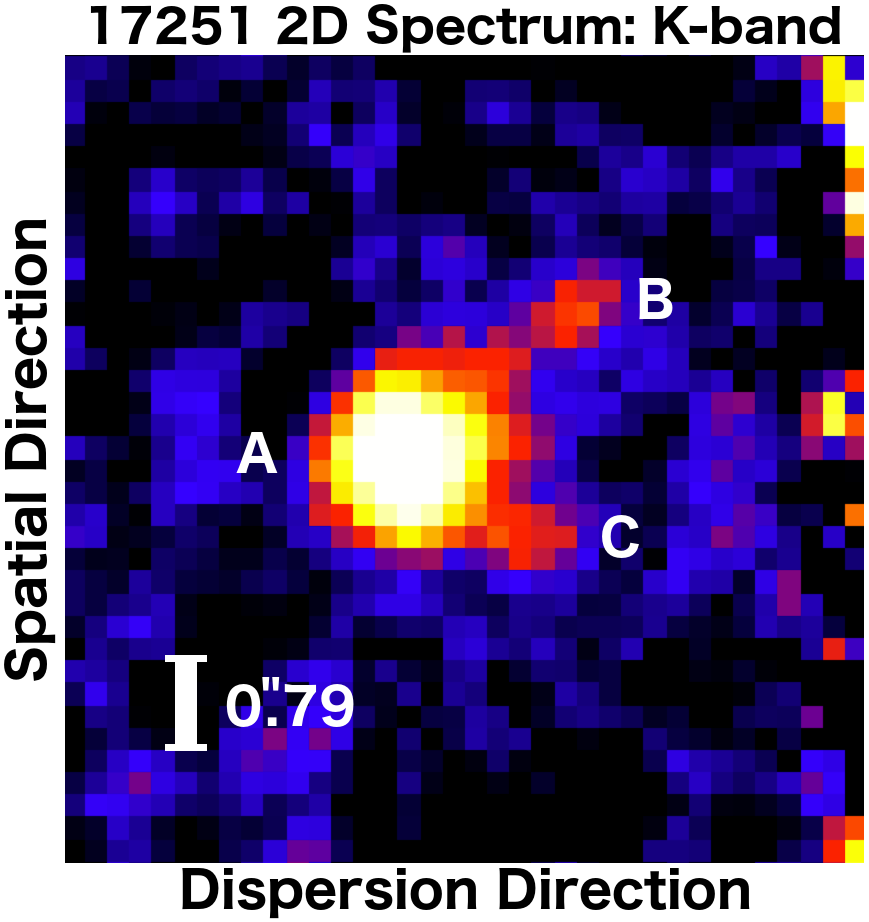}
    \caption{A cutout of the region in the 2D MOSFIRE, K-band spectrum
      of galaxy 17251 smoothed using a boxcar kernel with a width of 1
      pixel. This galaxy is found to have significant
      velocity structure with emission in regions B and C estimated to have
      velocity offsets of +52 and +38 km s$^{-1}$, respectively, from
      the primary galaxy denoted as region A. Both regions B and C
      have sizes comparable to the 0$\farcs$79 seeing of our
      observations (indicated by the white scale bar), meaning they are
      only marginally resolved. This complex structure may be indicative
      of an ongoing merger, which could account for both the large
      value of R23 (43.8) and the small value of O32 (0.2), as [OII]
      ($\lambda\lambda$3727 {\AA}) emission is
      known to be enhanced by shocks in regions between merging
      galaxies \citep[e.g.][]{epinat18}.}
    \label{fig:17251}
\end{figure}

Galaxy 17251 is plotted with an open, purple circle was found to have
an extremely large R23 value of 43.8, which is roughly 5 times higher
than the average value for galaxies in our sample with reliable
H$\beta$ detections. This large value is also at odds with theoretical
expectations from photoionization modeling, which give an upper bound of $\sim$10
\citep{kewley02,nagao06}. Examining the 2D spectra from MOSFIRE, we
find significant velocity structure in the [OIII] ($\lambda$5007 {\AA}) emission
line as shown in the boxcar smoothed (kernel width = 1 pixel) cutout in Figure \ref{fig:17251}. The primary
galaxy is located at emission peak A while peaks B and C are found to
have velocity offsets for the [OIII] ($\lambda$5007 {\AA}) emission line of +38 and +52 km
s$^{-1}$ respectively. We indicate the average seeing of our K-band
observations of 0$\farcs$79 with a white scale bar, a size comparable
to peaks B and C.  Thus, the size of these structures is consistent
with the seeing FWHM, thus, they are likely real but only marginally
resolved in our MOSFIRE observations.

This complex structure may be indicative of an
ongoing merger, a possibility consistent with the irregular morphology seen
in the HST image shown in Figure \ref{fig:HST}. We reextracted the
spectra in narrow spatial apertures in an attempt to capture the
properties of the individual emission line regions. We find that
region A and C have similar values of [OIII]/H$\beta$ of 9.6 and 7.8,
respectively. This may be due to the fact that there is
significant spatial overlap between these regions in the 2D spectrum
making it difficult to extract a spectrum for region C with no contribution
from region A. Region B is found to exhibit no
measurable H$\beta$ emission and relatively weak [OIII] ($\lambda$5007
{\AA}) emission.

Shocks, which are common in ongoing galaxy mergers, are known to have strong effects on emission line
ratios \citep[e.g.][]{rich15}, and in particular can greatly enhance
[OII] ($\lambda\lambda$3727 {\AA}) emission relative to [OIII]
($\lambda$ 5007 {\AA}) and H$\beta$ \citep{epinat18}. This
could help to explain both the high R23 and the low O32 found for
this galaxy. The poorer seeing of our H-band observations
(0$\farcs$96 vs 0$\farcs$79 for the K-band) as well as contamination
from sky emission lines prevent us from
performing a similar extraction of spectra for individual emission line peaks for
[OII] ($\lambda$3727 {\AA}) emission to test this. Regardless, the
fact that shocks associated with merger activity can significantly
impact the observed line ratios is important in the comparison between
$f_{esc}(LyC)$ and O32 both here and in all O32 studies of LCEs.. This is particularly interesting as we
estimate $f_{esc}(LyC)=0.37$ for this galaxy from photometry. 

\subsection{Comparison with Photoionization Modeling}\label{section:mappings1}

\begin{figure}
  \includegraphics[width=\columnwidth]{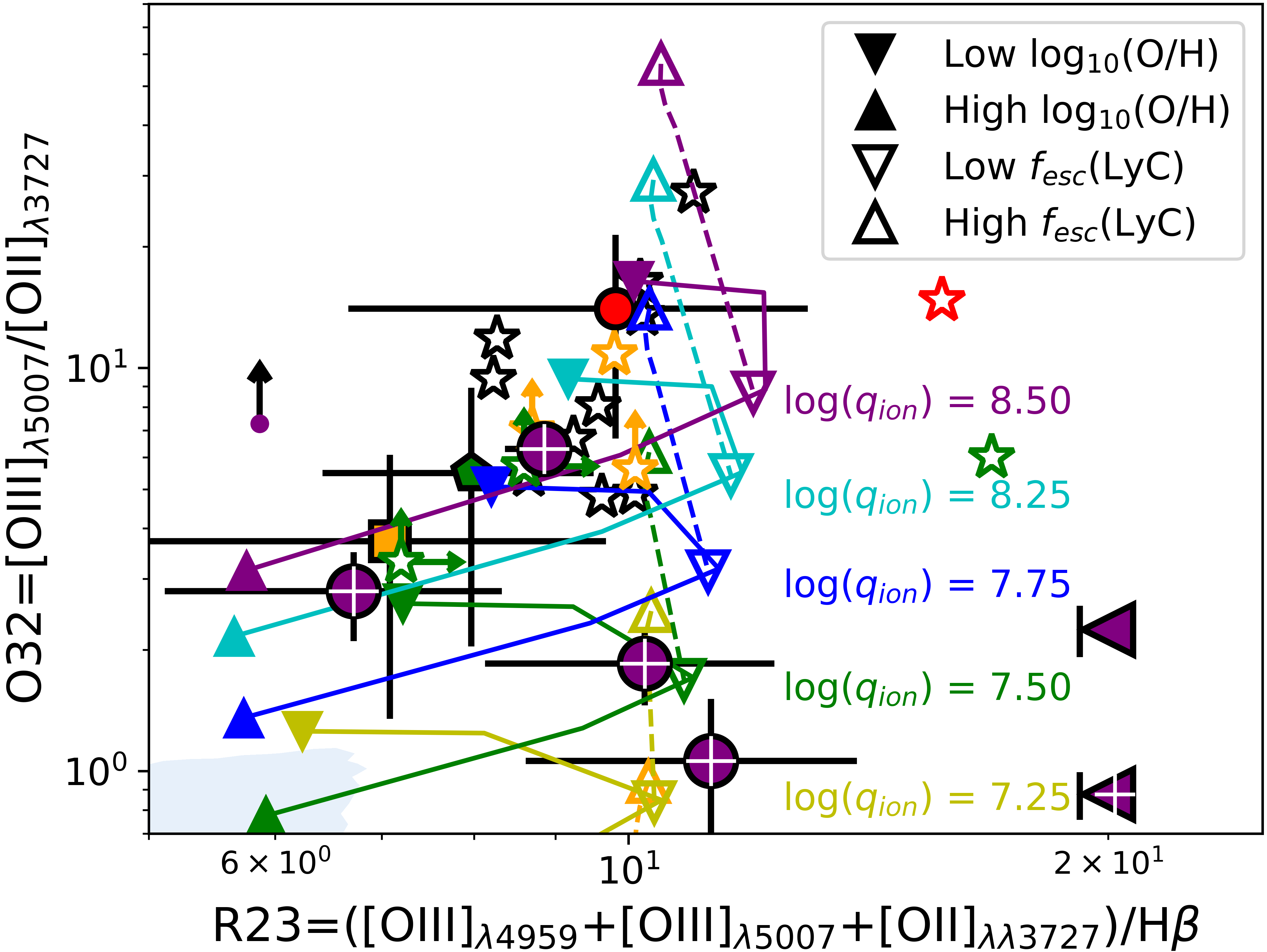}
  \caption{R23 vs O32 diagram, similar to Figure
    \ref{fig:R23O32_1}, however now we have zoomed into the
    upper-right region where highly ionized galaxies are found. With
    the exception of LCEs (white stars), individual galaxies
    in comparison samples have been removed for simplicity. Symbols
    common to both figures are the same as in Figure
    \ref{fig:R23O32_1}. The major addition in this Figure are
    photoionization modeling tracks output by the MAPPINGS V
    code. Solid lines show tracks for ionization bounded models of
    constant ionization parameter,
    $q_{ion}$ (values shown in the annotated text with corresponding
    colours), with increasing $12+log_{10}(O/H)$ as indicated by the filled
    triangles. Dashed lines show tracks with increasing
    $f_{esc}(LyC)$, as indicated by the open triangles,
    at fixed $q_{ion}$ and $12+log_{10}(O/H)$. These models clearly show the three-fold
    degeneracy between $q_{ion}$, $12+log_{10}(O/H)$, and $f_{esc}(LyC)$ in the observed
    R23-O32 space.}
  \label{fig:R23O32_2}
\end{figure}

Here we present the emission line ratios calculated in our
photoionization models, described in Section
\ref{section:pmod}. Separate metallicity and $f_{esc}(LyC)$ tracks in
R23 vs O32 at fixed ionization parameter, $q_{ion}$, are shown in Figure
\ref{fig:R23O32_2}. Solid lines represent ionization bounded models
($f_{esc}(LyC)$ = 0.0) at various metallicities, whereas dashed lines
represent density bounded models at a fixed metallicity from
$f_{esc}(LyC)=0.0$ to $f_{esc}(LyC)=1.0$. 
Tracks are shown over observed values for our sample,
as well as average values for comparison samples described in Section
\ref{section:linerats}.

\begin{figure*}
  \includegraphics[width=\textwidth]{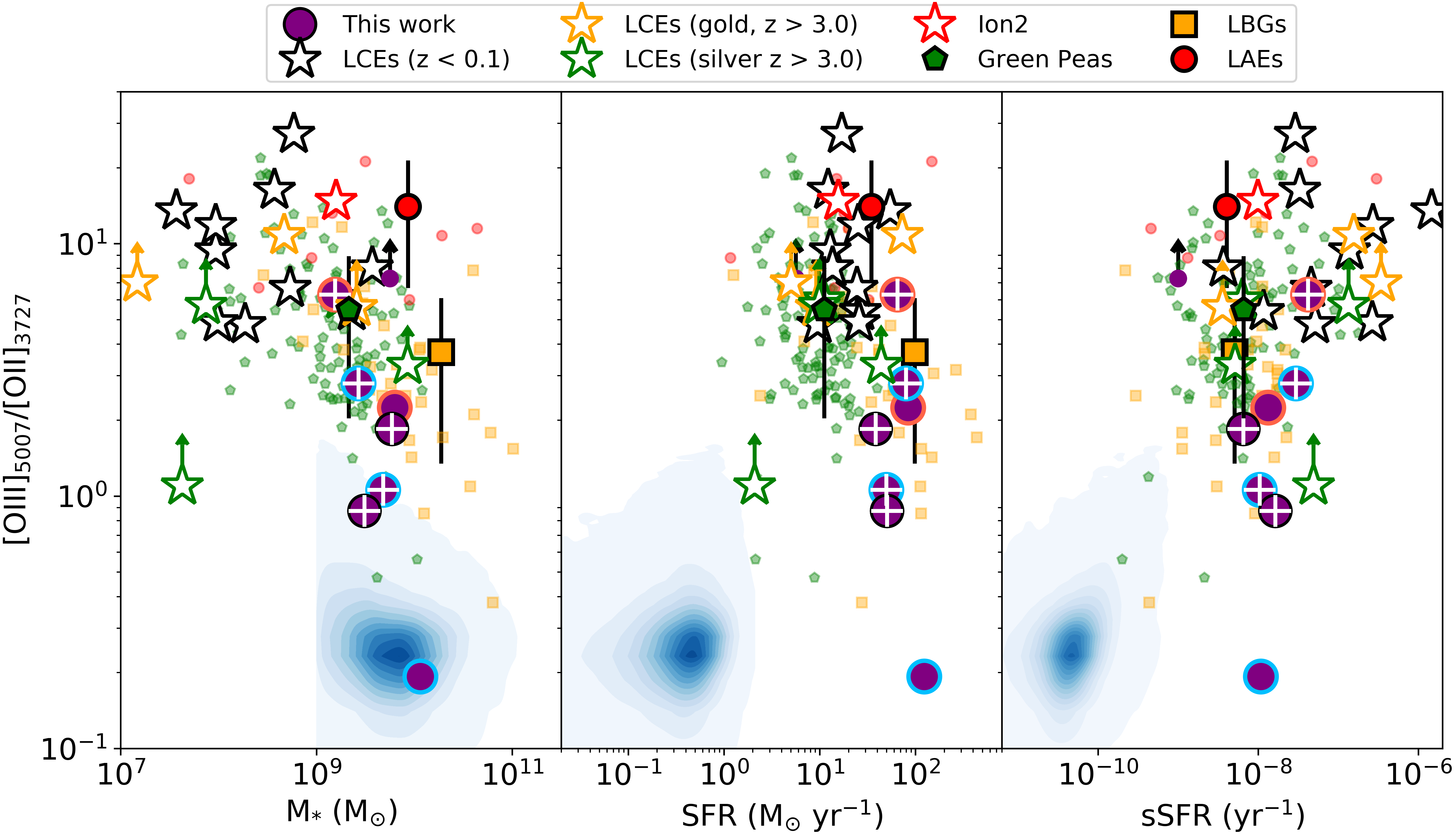}
  \caption{O32 as a function of $M_{*}$, SFR, and sSFR for galaxies in
  our MOSFIRE sample as well as low and high redshift comparison
  samples. Plotted symbols are the same as in Figure
  \ref{fig:R23O32_1}. Purple circles with blue borders indicate
  the three galaxies having $u$-band flux consistent with $f_{esc}(LyC) > 0.0$
  as estimated in Section \ref{section:fescphot}. The two purple
  circles with red borders indicate galaxies where upper limits on
  $u$-band flux provide non-zeros upper limits on $f_{esc}(LyC)$. The
  remaining three purple circles are galaxies where $f_{esc}(LyC)$ is
  consistent with zero even within the photometric uncertainty. 
  Galaxies in our sample have $M_{*}$ comparable
  to low mass LBGs and high mass green peas, though up to 1 dex more
  massive than confirmed high-$z$ LCEs. These results are likely
  reflective of relatively high stellar mass limits imposed for reliable
  photometric detections at high redshift (i.e. our desire to target
  bright galaxies). SFRs of our sample are comparable to LBGs, which
  are typically more star forming than green peas and LCEs. Finally,
  in terms of sSFR, galaxies in our sample are quite similar to
  confirmed LCEs.}
  \label{fig:pcomp}
\end{figure*}

Results presented for ionization bounded models in Figure
\ref{fig:R23O32_2} mirror those presented in \citet{nakajima14}. This
is not surprising as R23 vs O32 tracks from \citet{nakajima14} were
based on values from a previous version of the MAPPINGS code presented
in \citet{kewley02}. The key point illustrated by these ionization
bounded models is a clear degeneracy between metallicity and
$q_{ion}$. \citet{nakajima14} show that, by using an iterative process, an
estimate of gas-phase metallicity, $12+log_{10}(O/H)$, and $q_{ion}$
can be calculated based on the solid
curves in Figure \ref{fig:R23O32_2}. Following the example of
\citet{nakajima14}, we also 
calculate $q_{ion}$ and $12+log_{10}(O/H)$ finding values in the range 7.67
< $q_{ion}$ < 8.10 and 8.18 < $12+log_{10}(O/H)$ <
8.86. \citet{nakajima14} show that these are similar to the average
values for green peas and LBGs.

Also shown in Figure \ref{fig:R23O32_2} as dashed lines are tracks for
density bounded photoionization models calculated as described in
Section \ref{section:pmod}. Each track is calculated at a fixed
metallicity of $12+log_{10}(O/H)=8.48$ with $f_{esc}(LyC)$ increasing from
0 to 1.0. As expected, the $f_{esc}(LyC)=0.0$ model overlaps the
ionization bounded model at the corresponding $q_{ion}$ and metallicity,
while nonzero $f_{esc}(LyC)$ models extend to higher O32 with little
variation in R23. 

Ionization and density bounded models in Figure \ref{fig:R23O32_2}
demonstrate the threefold degeneracy between metallicity, ionization parameter, and
$f_{esc}(LyC)$ in the R23 vs O32 plane. Clearly, for LyC emitting
galaxies, an iterative approach in calculating ionization parameter and metallicity
based on R23 and O32 \citep[e.g.][]{nakajima14}, where an ionization
bounded model has been assumed may not be applicable. 

Of course, the
comparison here is between ionization and density bounded models with
simple geometry. In cases where LyC photons escape through holes in
geometrically complex H\,{\sc ii} regions, it is still unclear if
either of these approximations (density bounded versus ionization
bounded) can be applied to integrated line flux measurements
\citep[see also discussion in][]{fletcher18,steidel18}. As we have
noted, the majority of galaxies in our sample have O32 values
suggestive of a negligible $f_{esc}(LyC)$. Thus, the three low O32 galaxies with
accurate R23 values are estimated in the ranges 8.18 < $12+log_{10}(O/H)$ <
8.86 and 7.67 < $log_{10}(q_{ion})$ < 7.76. Although there is a clear
degeneracy between metallicity, $q_{ion}$, and $f_{esc}(LyC)$, there
is also evidence of an empirical trend between metallicity and
$q_{ion}$ \citep[e.g.][]{dopita06,sanders16,onodera16,kojima17} that
may alleviate some difficulty. We discuss theoretical and empirical
trends between $f_{esc}(LyC)$ and O32 further in Section
\ref{section:mappings2}. 

\section{Discussion}\label{section:discussion}

In this work, we have selected eight galaxies that are likely emitters of LyC at
$\langle z \rangle =3.17$ based
on their restframe UV colours indicative of LyC
emission (as detected in the $uS$ band, see Section 
\ref{section:colsel}), and explored their rest-frame optical emission line
properties using spectra from MOSFIRE H and K band observations. We
have shown in Sections \ref{section:linerats} that the O32 values found in our selected
sample are comparable to other samples of star-forming galaxies at
$z>3.0$, as well as highly ionized galaxies (including confirmed LCEs) at low redshift.

In this Section
we examine the integrated properties of galaxies in our sample,
including photometrically estimated $f_{esc}(LyC)$, in
comparison to samples of suspected (and confirmed) LyC emitters at
high and low redshift. This can give a better indication 
of the likelihood of LyC escape from these galaxies in lieu of
spectroscopic LyC detections. In the future, we will compare this analysis
with LyC detections (or upper limits) from our
LRIS observations and $f_{esc}(LyC)$ measurements from photometry and
spectroscopy of
higher redshift targets (Cooke et al. in preparation, Me\v{s}tri\'{c}
et al. in preparation). In Section
\ref{section:future} we discuss how lessons learned from this pilot study have
refined our selection methodology with the aim of improving our
efficiency of selecting LyC emitters from photometric data and
enabling an unbiased investigation of the full $z\sim3-4$ galaxy
population. Finally, in Section \ref{section:mappings2}
we discuss the implications of the density bounded photoionization
models from \ref{section:mappings1} in comparison with the recent
empirical relationship between O32 and $f_{esc}(LyC)$. 

\subsection{Do Photometric LCE Candidates Represent Key Contributors to Reionization?}\label{section:msfr}

In Figure \ref{fig:pcomp} we explore the relationship between the O32
ratio and $M_{*}$, SFR, and sSFR for galaxies in our
sample as well as those in comparison samples shown in Figure
\ref{fig:R23O32_1}. Green pea galaxies from \citet{yang16} and
\citet{cardamone09} and LCE galaxies from \citet{izotov16} and
\citet{izotov18b} have had their stellar masses and SFRs converted from the
\citet{salpeter55} IMF used in these studies to the \citet{chabrier03}
IMF used in this work using conversion factors found in the literature
\citep[e.g.][]{kennicutt12,speagle14}. These conversions are largely
negligible, however, and have no appreciable effect on our
results. Stellar masses and SFRs of high 
redshift comparison samples all adopt a similar IMF to ours, thus
requiring no adjustment. 

Figure \ref{fig:pcomp} provides a rough assessment of the
likelihood that our selected galaxies are
significant emitters of LyC photons. Compared with other samples shown
in Figure \ref{fig:pcomp}, galaxies 
in our sample are relatively uniform in $M_{*}$, but exhibit a wide
spread in the value of O32. In terms of both $M_{*}$
and SFR, our sample is most similar to intermediate 
mass green pea galaxies and low mass, but bright, LBGs. We note that here $M_{*}$
comes from SED fitting where nebular emission lines have been included in the fit as it
has been shown that failing to do so results in an overestimate of
$M_{*}$ \citep[e.g.][]{forrest17}

The key comparison with our sample in Figure \ref{fig:pcomp} is with
confirmed LCEs. Considering $M_{*}$, 12/16 LCEs are 
found th have masses 1-2 orders of magnitude less than
the majority of galaxies in our sample. This is consistent with hydrodynamical
simulations that find $f_{esc}(LyC)$ significantly decreases with
increasing stellar mass \citep{yajima11,wise14}. The most massive LCEs
from \citet{izotov18}, \citet{izotov18b}, and \citet{fletcher18} have
comparable mass to our sample and overlap in O32 with our highest O32
galaxies. Furthermore, the SFRs of galaxies in our sample overlap with
the high SFR end of LCEs. Assuming these galaxies form stars
from a similar initial mass function, this would imply comparable
numbers of massive stars and, thus, a similar production rate of
ionizing photons. The caveat here is that, as we estimated in Section
\ref{section:fescphot}, galaxies in our sample may in fact have a low
efficiency of ionizing photon production, $\xi$ (although our estimate
is highly uncertain due to unknown effects of slit-loss). 
Considering sSFR we find again that galaxies in our sample are
comparable to those found for LCEs.
We note again that hydrodynamical simulations find a
correlation between increasing sSFR and increasing
$f_{esc}(LyC)$ \citep{yajima11,wise14}. 

We recall the fact that 5/8 of the galaxies in our
sample exhibit confirmed Ly$\alpha$ emission from our LRIS
observations. Similarly, the vast majority currently confirmed detections of LyC emission from spectra occur in
galaxies also exhibiting Ly$\alpha$ emission
\citep{heckman11,leitet13,leitherer16,debarros16,izotov16,izotov18b,izotov18,shapley16,vanzella16,vanzella17,fletcher18,steidel18}.
We
note that the Ly$\alpha$ properties of the redshift
$\sim$2.5 LCE from \citet{bian17} are unknown
as this galaxy lacks spectroscopic observations targeting the
Ly$\alpha$ wavelength range. Regardless, there
appears to be a connection between Ly$\alpha$ and LyC escape with
\citet{verhamme17} showing that $f_{esc}(Ly\alpha)$ is correlated with
$f_{esc}(LyC)$ at low redshift. Furthermore, \citet{izotov18} and \citet{izotov18b} show a
correlation between $f_{esc}(LyC)$ and the separation of red and blue
peaks of the double peaked Ly$\alpha$ line in low-redshift LCEs. 
Interestingly, our second strongest LCE, galaxy 17800 with
$f_{esc}(Lyc) = 0.21\pm0.09$ exhibits Ly$\alpha$ in
absorption. Additionally, galaxy 12676, which we estimate a non-zero
upper limit on $f_{esc}(LyC)$ of 0.04 is also a Ly$\alpha$ absorber. For
the remaining 3 LCEs in our sample, the resolution of our LRIS observations is not
sufficient to resolve any possible double peak of the observed Ly$\alpha$
emission lines and thereby test the prediction of \citet{verhamme17}
at high redshift.

O32 measurements for our sample allow us to assess the applicability
of an O32-$f_{esc}(LyC)$ relation to our sample. The 3 galaxies in our
sample that have nonzero estimates of $f_{esc}(LyC)$
based on CLAUDS $u$ and $r'$ band photometry and the two additional
galaxies with non-zero upper limits on $f_{esc}(LyC)$ (see Section
\ref{section:fescphot}) are indicated in Figure \ref{fig:pcomp} using
coloured edges. Symbols with blue edges are firm detections while
those with red edges indicate upper limits based on the photometric
uncertainty of the CLAUDS $u$-band observations. Though not detected
in [OII] ($\lambda\lambda$3727 {\AA}), the lower limit of O32 for
galaxy 13459 of 8.5 is larger than any other galaxy in our sample, and
this galaxy has $u$ band flux consistent with $f_{esc}(LyC)=0.0$. The
highest O32 observation for [OII] ($\lambda\lambda$3727 {\AA})
detected galaxies is galaxy 16067, for which we estimate a relatively
low upper limit on $f_{esc}(LyC)$ of 0.06. We reiterate here that
the [OII] ($\lambda\lambda$3727 {\AA}) flux may be overestimated due
to OH contamination (see
Section \ref{section:elf}), thus its true O32 may be even
higher. Among the three galaxies 
with firm non-zero measurements of $f_{esc}(LyC)$, only galaxy 17800
has an O32 value significantly higher than 1. Most interestingly,
galaxy 17251 has both the highest estimate of
$f_{esc}(LyC)=0.37\pm0.08$ and the lowest O32 of 0.19$\pm$0.06.
If the measured $f_{esc}(LyC)$ values for our sample are confirmed
spectroscopically, this suggests a high O32 is not an absolute
requirement for LyC escape from star-forming galaxies.

Taking a closer look at the three low O32 LCEs in our sample, we see that
two, 17251 and 17800, have some evidence of merger
activity indicated by their complex HST morphologies seen in Figure
\ref{fig:HST}. Galaxy 17251, in particular, which has an extremely low
O32, exhibits complex [OIII] ($\lambda$5007 {\AA}) emission in our 2D
spectrum (see Section \ref{section:17251}). It has been shown that the
presence of shocks, common in ongoing mergers, can significantly
enhance [OII] ($\lambda\lambda$3727 {\AA}) relative to [OIII]
($\lambda$5007 {\AA}), thus reducing the observed O32
\citep{rich15,epinat18}. These two high $f_{esc}(LyC)$ galaxies
highlight the possibility of a significant role of
galaxy mergers in the EoR \citep[see also][]{bergvall13}.  We discuss
further the comparison between O32 
and our photometric estimates of $f_{esc}(LyC)$ below in Section
\ref{section:mappings2}.

The emerging picture of the driver of the EoR is that low mass
($\lesssim$10$^{8}$ M$_{\odot}$), low luminosity galaxies with high sSFR
($\gtrsim$1 Gyr$^{-1}$) may be the key sources of reionizing photons
\citep{wise09,yajima11,wise14,bouwens15,paardekooper15}. This may
relate to the fact that low mass galaxies are smaller in size
\citep[e.g.][]{lange15,bouwens17} meaning star-forming regions are more likely
to be located near the edge of a galaxy where mechanical
star-formation feedback (i.e. supernova blast waves) can more easily
reach the IGM. Furthermore, low mass galaxies host less interstellar
dust \citep{garn10}, meaning LyC emission escaping H\,{\sc ii} regions is less
likely to be attenuated by dust before reaching the IGM in lower mass
galaxies. 

In this context, given the relatively low $f_{esc}(LyC)$
estimates for the majority of our sample from CLAUDS $u$-band
photometry (which are bright and typically more massive than other
known LCEs), we expect
that galaxies similar to those selected in this work are not likely to
be the key contributors of ionizing radiation during the EoR. Regardless,
galaxies such as these will be relevant in estimating the overall ionizing photon
budget at $z>6$. Furthermore, we reiterate that we have estimated a
relatively low, though uncertain, value of $\xi$ for our sample (see
Section \ref{section:fescphot}). If this is reflective of a lower
value of $(L_{\nu}(LyC)/L_{\nu}(UV))_{int}$,  the efficacy
of such galaxies to contribute significantly to reionizaion would be
further reduced. Finally,
it should be stressed that photometric $f_{esc}(LyC)$ estimates
presented here are tentative values as LyC photon detections in our
LRIS spectra for individual galaxies are still being assessed.

\subsection{Improving Selections of High Redshift LyC Emitting Galaxies}\label{section:future}

The results of this pilot survey show that the employed selection
criteria have not provided a clean sample of $z\simeq3$ LCEs. In
particular, the redshift range probed results in a significant and
difficult to distinguish contribution to the $u$ band photometry from
non-LyC photons. This complication means that $f_{esc}(LyC)$ estimates
from photometric observations are model dependent, thus prevents us
from definitively identifying galaxies in our sample as LCEs. 

We also draw attention to the HST imaging for our sample
presented in Figure \ref{fig:HST}. As we pointed out in Section
\ref{section:colsel}, HST morphologies show elongation and/or multiple
continuum peaks. These morphological signatures are indicative either
of ongoing merger activity, multiple star-forming regions, or line-of-sight contamination by
unrelated, low redshift galaxies. The latter
possibility has been shown to prevent a reliable LyC detection
\citep[e.g.][]{vanzella10,vanzella12} as such low redshift contamination
would have a strong impact on our $uS$ magnitudes. In our sample, no emission
lines from lower redshift galaxies
are apparent in either our MOSFIRE or LRIS spectra, meaning that our
O32 measurements are likely unaffected by any possible low redshift
contamination. The one caveat here is that the lack of low redshift
emission lines in our MOSFIRE spectrum does not exclude low redshift,
non-star-forming companions that may affect observations of LyC
emission. Regardless of the issue of low redshift contamination, our results have
shown that ongoing mergers may in fact exhibit heightened LyC escape,
thus galaxies with disturbed morphologies should not be avoided
entirely.

The challenges and results of this pilot survey
have informed and improved our future sample selection of LCE
candidates. One goal of our future selection criteria is to avoid
both galaxies with $\lambda$ $>$ 912 {\AA} emission in $u$-band
photometry and galaxies having any possible evidence of low redshift
contaminants based on high-resolution HST imaging, though disturbed
galaxies indicating possible ongoing mergers are not excluded. This will be achieved by
including two additional considerations to our sample
selection to improve the efficiency of building our sample of LCEs:
\begin{itemize}
  \item First, we give higher priority to galaxies at $z>3.4$ with $u$
    band detections with future selections using the CLAUDS $u$ band filter
    which exhibits a sharp cutoff at 4000 {\AA}, corresponding to 912
    {\AA} at $z=3.4$. We have supplemented this sample with galaxies
    selected in regions with HST F336W imaging (GO 15100; PI Cooke), which cleanly
    probes LyC emission down to $z=3.05$. 
  \item Second, we impose an initial preference for compact galaxies
    with single UV continuum peaks, and ``clean'' detections: those
    galaxies with no other objects within 1$\farcs$0 of the primary
    source. After this, a secondary preference will be given to
    galaxies with disturbed morphologies indicative of on-going
    mergers.
\end{itemize}
The first criterion allows us to be confident that all photons
from $z>3.4$ galaxies detected in the $u$ band are from LyC emission rather 
than longer wavelength Ly$\alpha$ forest light. This higher redshift
cutoff will also aid with spectroscopic detection of LyC photons as the
LyC region moves to more sensitive portions of the LRIS CCD.
The second criterion ensures we will primarily select galaxies that are the least
likely to be affected by lower redshift, line-of-sight
interlopers (at the cost of not exploring extended or clumpy galaxies or galaxy
mergers). Furthermore, the requirement of compactness follows the work
of \citet{izotov16} who find 
that at low redshift, LCEs are found to be among the most compact
galaxies from their parent samples of green peas. On the other hand, our secondary
preference for galaxies with disturbed HST morphologies allows us to
probe possible galaxy mergers, which we have found in this work to be
efficient emitters of LyC radiation. Due to the low covering fraction
of primary, compact targets, the inclusion of secondary, disturbed
targets further improves the efficiency of our observational
strategy.

\begin{figure*}
  \includegraphics[width=\textwidth]{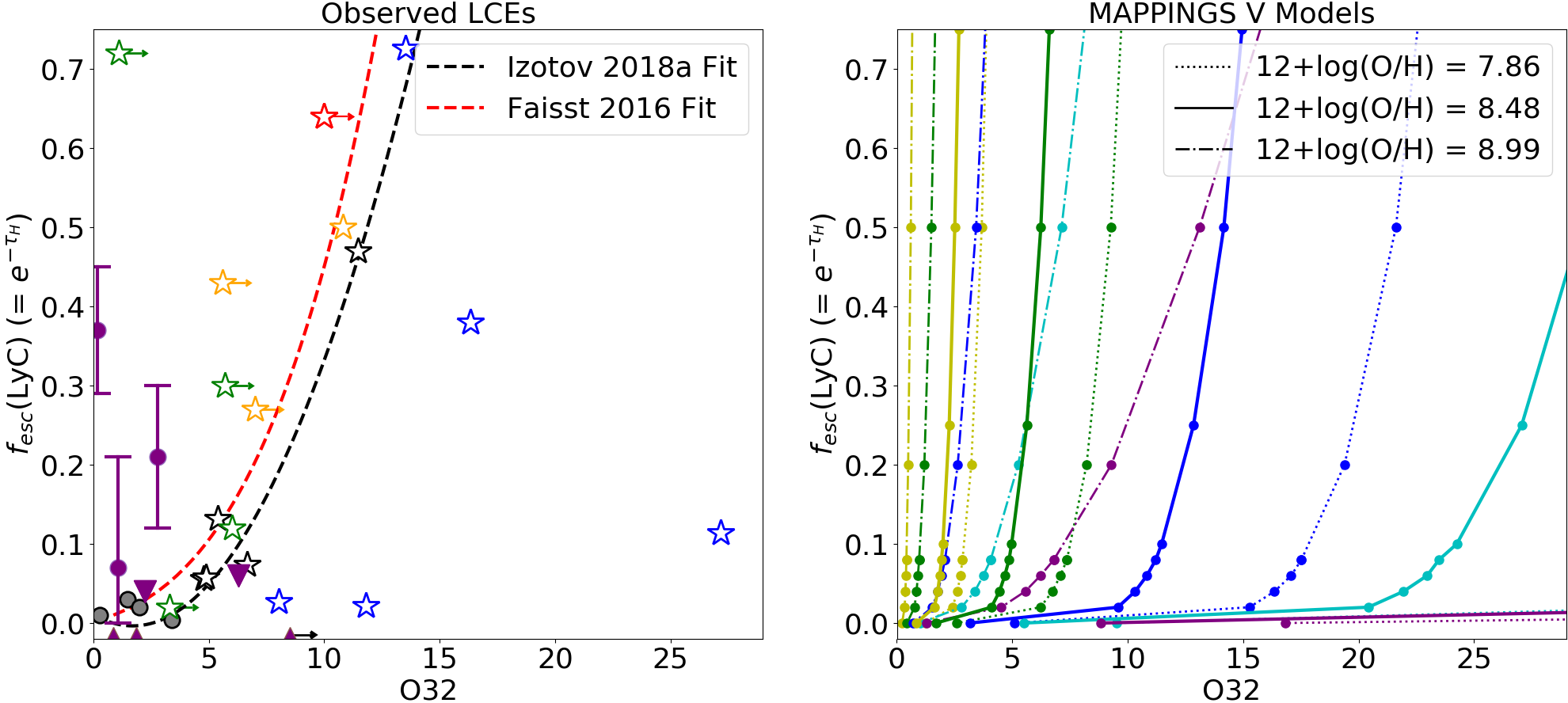}
  \caption{A comparison of the empirical relations between O32 and
    $f_{esc}(LyC)$ suggested by \citet{izotov18} and \citet{faisst16} (left) with the theoretical
    predictions calculated using MAPPINGS V (right). On the left, LCEs at $z
    \simeq 0.05$ from \citet{izotov16} and \citet{izotov18} are shown as black, open stars as in Figures
    \ref{fig:R23O32_1} and LCEs from \citet{izotov18b} are shown with
    blue, open stars. Lower redshift LCEs are indicated with grey
    circles. The open red star indicates the high redshift
    LCE, Ion2 \citep{debarros16} and orange and green open stars show
    ``gold'' and ``silver'' subsamples of LCEs from the LACES survey
    \citep{fletcher18}. Observations for our sample are
    shown with purple symbols. Circles with error bars show galaxies
    with nonzero $f_{esc}(LyC)$ from CLAUDS $u$-band photometry,
    downward triangles show upper limits on $f_{esc}(LyC)$ based on
    CLAUDS $u$-band uncertainty, and small, upwards triangles show
    galaxies with $f_{esc}(LyC)=0.0$ even when uncertainty in $u$ is
    considered. The empirical fits presented in
    \citet{izotov18} and \citet{faisst16} and shown by black and red
    dashed lines, respectively. The fit of \citet{faisst16} includes
    Ion2 and all black LCEs except for the one at the highest O32 and
    the fit of \citet{izotov18} excludes Ion2 and includes all black,
    open stars. Blue open stars were reported more recently than
    either fit, and thus were not included. On the right, coloured lines show the outputs of the MAPPINGS V 
    code at fixed $Z$ with colours indicating different $q_{ion}$, identical
  to Figure \ref{fig:R23O32_2}. Different line styles for MAPPINGS V
  results correspond to different $12+log_{10}(O/H)$ as indicated in the
  legend.}
  \label{fig:fesctracks}
\end{figure*}

These
selection criteria have been implemented in recent and upcoming
observations with LRIS and HST targeting LyC emission, which can
provide final confirmation of our selection methodology that has no
bias on color selection or restframe optical emission line strengths.
A more complete description of our
improved selection criteria and preliminary results of recent
HST observations based on this selection will be presented in Cooke at
al. (in preparation).

\subsection{The Correlation Between $f_{esc}(LyC)$ and O32}\label{section:mappings2}

In this Section we discuss the recent suggestion by \citet{faisst16} and
\citet{izotov18} of an empirical correlation between O32 and
$f_{esc}(LyC)$ for galaxies with detected LyC emission. Here we
comment both on the predicted $f_{esc}(LyC)$ for our galaxy sample
based on the observed O32 values as well as the comparison with
density bounded models computed using the MAPPINGS V code as described
in Section \ref{section:pmod}. 

The left panel of Figure \ref{fig:fesctracks} compares the observed distribution of
confirmed LCEs and the empirical fits in O32 vs $f_{esc}(LyC)$ space
while the right panel shows the theoretical predictions of our MAPPINGS
V density bounded
H\,{\sc ii} region models. Our fixed metallicity models were calculated on
grids with 9 values of 
decreasing HI optical depth, $\tau_{HI}$, simulating $f_{esc}(LyC)$
values from $\sim$0.999 down to 0.001 and 8 values of $log_{10}(q_{ion})$ from 6.5
to 8.5. We calculated three sets of grids at fixed $12+log_{10}(O/H)$ of
7.86, 8.48, and 8.99 shown in the right side of Figure \ref{fig:fesctracks} as dotted,
dot-dashed, and solid lines, respectively with colours indicating
$log_{10}(q_{ion})$ matched to the values shown in Figure
\ref{fig:R23O32_2}. 

At low
$12+log_{10}(O/H)$ and fixed $q_{ion}$ we find that the largest variation in O32 with
$f_{esc}(LyC)$ occurs below $f_{esc}(LyC)$ = 0.10. For low metallicity
models there is still some increase in O32 with increasing
$f_{esc}(LyC)$ above $f_{esc}(LyC)$ = 0.10, with larger variation at
higher ionization. Consider the $12+log_{10}(O/H)=8.48$, $log_{10}(q_{ion})=7.75$ model
(solid blue line) for instance, from $f_{esc}(LyC)=$0.01 to
$f_{esc}(LyC)=$0.10, we find an increase in O32 from 3.2 to 11.5. For
the same model, however we find between $f_{esc}(LyC)=$0.10 to
$f_{esc}(LyC)$=0.75 only a comparatively modest increase in O32 from
11.5 to 14.9.
Although we find a more constantly varying $f_{esc}(LyC)$
as a function of O32 for high metallicity and high ionization models
(e.g. the purple dot-dashed line in Figure \ref{fig:fesctracks})
examples of highly ionized, high metallicity galaxies are not present
in the literature \citep[e.g.][]{kojima17}.

Thus, our MAPPINGS V
density bounded H\,{\sc ii} region models suggest that at low metallicity, O32
may have little constraining power on the escape fraction of LyC above
$f_{esc}(LyC)$ > 0.10. At high metallicities, O32 may be more
predictive of $f_{esc}(LyC)$ with the caveat the MAPPINGS V models are
one-dimensional. Figure \ref{fig:fesctracks} clearly shows
that an accurate measurement of both metallicity and ionization
parameter will be essential in reliably constraining  $f_{esc}(LyC)$
from optical emission line ratios alone. Empirically, there is strong
evidence that metallicity and ionization parameter are generally
correlated \citep{dopita06,sanders16,onodera16,kojima17} and such
observations will be vital in obtaining estimates 
of $f_{esc}(LyC)$ from strong line ratios.  Although our MAPPINGS V
models presented here, given full freedom in $q_{ion}$ and
$12+log_{10}(O/H)$,  can populate the entire O32 vs $f_{esc}(LyC)$
space (thus matching all known LCEs), these empirical results show
that certain combinations of metallicity and ionization may be
unrealistic. 

Low redshift LCEs from
\citet{izotov16} and \citet{izotov18} are shown by black open stars
while more recent observations from \citet{izotov18b} are shown with
blue open stars. Other low redshift LCEs, with $f_{esc}(LyC)$ measured by
\citet{chisholm17}, are shown with grey circles
\citep{borthakur14,alexandroff15,leitherer16}. Also shown are 
$z>3.0$ LCEs with red, orange and green stars representing Ion2
\citep{debarros16,vanzella16} and ``gold'' and ``silver'' subsamples
from the LACES survey \citep{fletcher18}, respectively. $f_{esc}(LyC)$
measurements from \citet{fletcher18} are achieved through SED fitting
where $(L_{\nu}(LyC)/L_{\nu}(UV))_{int}$ can vary from galaxy to
galaxy while other LCEs have $f_{esc}(LyC)$ computed assuming a fixed
value of 0.33. This should be kept in mind for such a direct
comparison as that presented in Figure \ref{fig:fesctracks} as the exact
$f_{esc}(LyC)$ likely depend on the computational method.

The empirical fits to LCEs of \citet{faisst16} and \citet{izotov18},
are shown in the left panel of Figure \ref{fig:fesctracks} using 
red and black dashed lines, respectively.
We note here that Ion2 was not included in the fit of
\citet{izotov18} while it was included in the fit of
\citet{faisst16} (assuming an lower limit of 0.50 on
$f_{esc}(LyC)$). LACES LCEs, as well as the blue, open stars
(representing low redshift LCEs of \citet{izotov18b})
were reported prior to \citet{faisst16} and \citet{izotov18}, thus
these data are not included in either fit. Interestingly, both empirical fits are 
qualitatively similar to the MAPPINGS model track for H\,{\sc ii} regions
with $12+log_{10}(O/H)=8.99$ and $log_{10}(q_{ion})=8.5$ (purple
dot-dashed line in right panel of Figure \ref{fig:fesctracks}). Galaxies from
\citet{izotov16} and \citet{izotov18} have measured
$12+log_{10}(O/H)$ between 7.62 and 8.00, suggesting that this
qualitative similarity is merely coincidental. Furthermore, as we have
mentioned, the combination of high ionization and high metallicity
represented by this track is unlikely for real galaxies.

The most striking discrepancy seen in the left panel of Figure \ref{fig:fesctracks} is
that 4/5 of the most recently reported $z\simeq0.05$ galaxies from
\citet{izotov18b} have O32 and $f_{esc}(LyC)$ values clearly in
disagreement with the earlier fits presented by \citet{faisst16} and
\citet{izotov18}. This discrepancy, which also has been commented on
in \citet{izotov18b}, casts serious doubt on a simple relationship
between O32 and $f_{esc}(LyC)$. A similar conclusion has also recently been
drawn by \citet{naidu18} who find a conspicuous lack of LyC emission
from stacked $u$ band photometry of 73 galaxies at $z=3.42-3.57$ that
are inferred to have O32$\gtrsim$4.

\begin{figure*}
  \includegraphics[width=\textwidth]{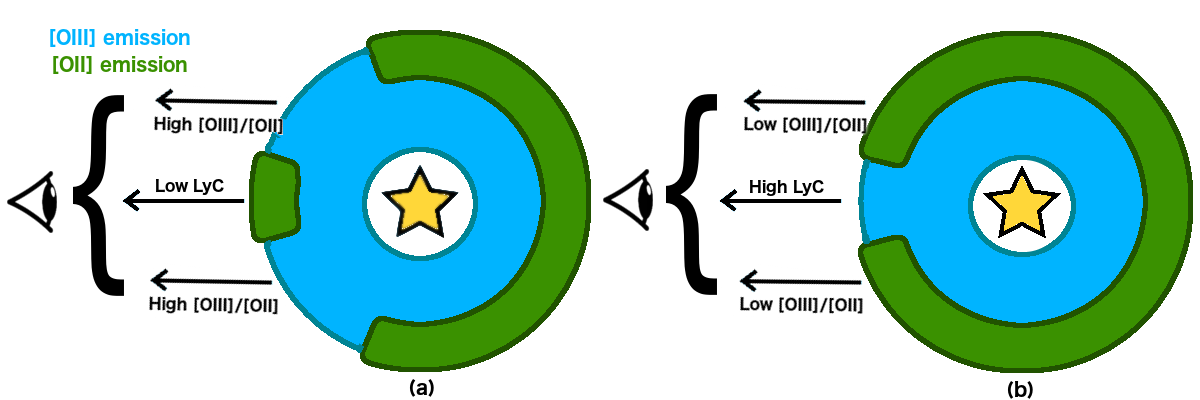}
  \caption{A conceptual picture of the differing geometric effects on
    observed LyC and scattered line emission from H\,{\sc ii} regions with
    holes. Blue indicates the hot, diffuse, [OIII] ($\lambda$5007 {\AA})
    emitting inner regions while green indicates cool, dense, [OII]
    ($\lambda\lambda$3727 {\AA}) emitting outer
    regions. \citet{giammanco05} and \citet{pellegrini12}, as well as
    our own MAPPINGS V analysis, have shown that a high gas density
    results in a lowered observed O32 value. \textit{(a)}: A low overall covering fraction
    of dense gas with the chance alignment of a small, dense clump
    that obscures the ionizing photon source(s). This would result in low
    $f_{esc}(LyC)$ and high O32 \citep[e.g.][]{izotov18b}. \textit{b}:
    A high overall covering fraction of dense gas with the chance
    alignment of a small hole in front of the source(s) of ionizing
    photons. This would result in a high $f_{esc}(LyC)$ and a low O32
    (e.g. our sample). See \citet{reddy16}, \citet{fletcher18}, and
    \citet{steidel18} for further discussion on the implication of
    clumpy geometry on $f_{esc}(LyC)$.}
  \label{fig:fescdiag}
\end{figure*}

One possible way to ease the
tension in such results is suggested by the MAPPINGS V tracks
presented in the right panel of Figure \ref{fig:fesctracks}. A more
complex relationship between
$12+log_{10}(O/H)$, $q_{ion}$, and $f_{esc}(LyC)$ may be required
\citep[e.g. by including known correlations between metallicity and
ionization parameter,][]{kojima17}.
Confirming this will require a larger sample of LCEs with
reliably measured metallicities. Regardless, observations of O32 in
confirmed LCEs suggest that all galaxies with
$f_{esc}(LyC)$ $>$ 0.20 also have O32 $>$ 10.0. Recent
observations of \citet{izotov18b} and lower limits on O32 for galaxy
13459 presented here (not detected at LyC wavelengths) clearly show
that a large O32 does not require a large observed value of $f_{esc}(LyC)$.

Another possible reason for a discrepancy between O32 vs
$f_{esc}(LyC)$ for \citet{izotov18b} galaxies and
the empirical relationships presented in \citet{faisst16} and \citet{izotov18} could be the lack of complex geometry
in our photoionization modeling. Observations of individual nearby H\,{\sc ii} regions
show them to be geometrically complex, with significant spatial variation in oxygen
line ratios indicative of localised, highly ionized regions from which
LyC could escape
\citep[e.g.][]{zastrow11,weilbacher15,kehrig16,keenan17,micheva18}. Similar regions in high redshift H\,{\sc ii} regions could
represent ``holes'' through which LyC flux could escape while other
areas of the H\,{\sc ii} region remain completely opaque to high energy
radiation \citep[e.g. models
of][]{zackrisson13,verhamme15,zackrisson16}. This also highlights a
limitation of our density bounded MAPPINGS V
photoionization models as these assume homogeneity in $\tau_{HI}$, thus a constant
$f_{esc}(LyC)$ across the entire nebula. Further discussion of
geometric effects on the inferred value of $f_{esc}(LyC)$ can be found
in recent works by \citet{steidel18} and \citet{fletcher18}. In
particular, \citet{fletcher18} point out that the stacking analysis of
strong [OIII] ($\lambda$5007 {\AA}) emitters employed by
\citet{naidu18} may not provide a reliable average $f_{esc}(LyC)$
where LyC photons escape from H\,{\sc ii} regions with complex
geometry. 

To illustrate the geometric effects on observable quantities, we focus on recent
observations of LCEs from \citet{izotov18b} with extremely high O32
and relatively low $f_{esc}(LyC)$. These observations may be the direct result of
differences in geometric dependency as LyC
photons must be emitted directly into our line-of-sight while line
emission can be scattered into our sight line. High O32 and low $f_{esc}(LyC)$ cases
could be explained by a nebular geometry in which dense gas clouds are
situated in our line of sight towards the ionizing source, giving a
low \textit{observed} $f_{esc}(LyC)$, while a hole (or multiple holes) through the nebula
away from our line of sight scatter significant amounts of [OIII] line
emission towards us. This geometric arrangement is illustrated
conceptually on the left side of Figure \ref{fig:fescdiag}. Such a case may also highlight an important
difference between the \textit{intrinsic} and \textit{observed} values
of $f_{esc}(LyC)$, as the latter quantity is highly dependent on
$\tau_{HI}$ in our direct sight line towards sources of ionizing
radiation. This may also suggest that, although recent evidence shows
a weak dependence of observed $f_{esc}(LyC)$ and O32, the O32 ratio
may indeed correlate with intrinsic $f_{esc}(LyC)$.

Testing this requires
full three-dimensional (3D) photoionization with radiative transfer and
could be performed using existing tools
\citep[e.g. Mocassin,][]{ercolano02}. Computing even a single
modelled H\,{\sc ii} region in 3D is computationally expensive, thus
exploring the parameter space represented by a suite of such models is
beyond the scope of the current work. We also note that, particularly
in massive galaxies, measured values are marginalised over multiple
H\,{\sc ii} regions within a given galaxy. Exploring 3D models of realistic
H\,{\sc ii} regions (and ensembles of these regions) in regards to LyC
escape, however, is an interesting avenue of future research. Given
the difficulties highlighted here with respect to O32 other
observational signatures expected to probe gas opacity and geometry in
H\,{\sc ii} regions should be explored further both theoretically and
observationally. These include high resolution observations of double
(or more) peaks in Ly$\alpha$ emission line profiles
\citep[e.g.][]{verhamme15,vanzella18}, rest-frame UV absorption line
optical depths \citep[probing neutral hydrogen covering fractions,
e.g.][]{steidel10,steidel18,jones13,leethochawalit16}, a comparison between
H$\beta$ EW and UV spectral slope \citep[providing a direct comparison
of photoionization and UV luminosity,
e.g.][]{zackrisson13,zackrisson16}, or the observation of an empirical
relation between $E(B-V)$ and neutral hydrogen covering fraction
\citep{reddy16}.  

Next we comment on the predicted $f_{esc}(LyC)$ for galaxies in
our sample based on their measured O32 values as well as the estimated
values from Section \ref{section:fescphot} based on CLAUDS $u$ band
photometry. The O32 and photometrically estimated $f_{esc}(LyC)$ values of our
galaxy sample are indicated in the left panel of Figure 
\ref{fig:fesctracks} using purple circles with error bars indicating
the spread attributable to error in the photometric measurement. Those
galaxies with $u$
band magnitudes consistent with $f_{esc}(LyC)=0.0$ are shown along the
x-axis using purple triangles. Photometric estimates of $f_{esc}(LyC)$
for 4/8 galaxies in our sample are found to be in agreement,
within photometric errors, with the fits of \citet{faisst16} and
\citet{izotov18}. Of the remaining four, our two highest O32 galaxies
(16067 and the lower limit for [OII] $\lambda\lambda$3727 {\AA} galaxy
13459) both fall
below the published relations, similar to \citet{izotov18b}, while the
remaining two are significantly above this relation. 

Galaxies falling above the relations of \citet{faisst16} and \citet{izotov18}
are currently only found in this study and the LACES survey
\citep{fletcher18}, and of particular interest. We note that from 
\citet{fletcher18} are not detected in [OII] ($\lambda\lambda$3727
{\AA}) thus O32 values are lower limits. If such a  population of low O32,
high $f_{esc}(LyC)$ galaxies
is confirmed at $z>3$, this could suggest an evolution in the ISM
properties of high redshift galaxies that more readily allow
ionizing photons to escape, even at low $q_{ion}$. A second
alternative is that this is due to a geometric effect in which a galaxy is viewed at a
particular line of sight directed at a hole in an H\,{\sc ii} region directly
in line with massive, ionizing stars while the overall covering
fraction of dense hydrogen in the nebula is quite high \citep[also
discussed in][]{fletcher18}. This peculiar
geometry is illustrated on the right side of Figure
\ref{fig:fescdiag}. We reiterate here
that a better understanding of the relationship between geometry and
$f_{esc}(LyC)$ in 3D, and how this relates to observed quantities in
2D, will likely shed light on the difficulties faced in uncovering
a reliable proxy for LyC escape during the EoR.

It is interesting that two of our galaxies falling above
the \citet{faisst16} and \citet{izotov18} relations in the left panel
of Figure \ref{fig:fesctracks}, 17251 and 17800, have merger-like HST
morphologies seen in Figure \ref{fig:HST}. Galaxy 17251 in particular,
with the lowest O32 in our sample, has additional evidence of being an
on-going merger from the complex [OIII] ($\lambda$5007 {\AA}) in our
2D spectrum (see Figure \ref{fig:17251}). The low O32 (lower limit)
LCE from \citet{fletcher18} has a low S/N in their HST imaging, thus
assessing the possibility of a merger from morphology is not
possible. As we have discussed in
Sections \ref{section:17251} and \ref{section:msfr}, shocks associated
with mergers can enhance [OII] ($\lambda\lambda$3727 {\AA}) relative
to [OIII] ($\lambda$5007 {\AA}), thus explaining to low observed O32
in these galaxies \citep{rich15,epinat18}. 
Observations of these
galaxies suggest that galaxy mergers may
significantly enhance $f_{esc}(LyC)$, but it is clear that predicting
$f_{esc}(LyC)$ for such objects from optical emission lines is
complicated due to the presence of shock excitation. Possible
mechanisms for an enhanced $f_{esc}(LyC)$ may include redistribution
of gas around existing stars or star-formation induced at the
outskirts of the merging system where supernova feedback can more
easily remove gas from around the newly formed stellar populations. 

Finally, we point
out that similar galaxies with low O32 and high $f_{esc}(LyC)$ are
not present in the low redshift samples of \citet{izotov18} and
\citet{izotov18b}. Green pea galaxies studied at high spatial
resolution and with integral field spectroscopy have shown the
population to be a mix of merging and non-merging systems
\citep{cardamone09,lofthouse17}, though the fraction of mergers in
green pea samples is still unclear due to low number
statistics. Regardless, the merger properties studied in
\citet{izotov18} and \citet{izotov18b} are not well constrained by
current observations. The high O32 for these samples suggests a lack
of shock excitation and thus may exclude on-going merger activity,
though this does not rule out the possibility they are recent merger
remnants. Retrograde mergers are known to funnel gas 
towards the galaxy centre resulting in compact and intense star
formation similar to that seen in green pea samples
\citep[e.g.][]{bassett17b}. Such a high star formation surface density
can enhance feedback and increase the likelihood of LyC escape.
A study of LyC emission from confirmed mergers and merger remnants
using HST spectroscopy at $z\simeq0.2-0.5$ would help to shed light on
the role of mergers in the escape of ionizing radiation.

\section{Summary and Conclusions}\label{section:snc}

The long-term goal of our observing program is to provide a method of
accurately estimating the level of LyC escape using rest-frame optical
nebular lines where direct LyC
detection is not available. An ideal calibration, which is explored in
this work, is based on the ratios of bright emission lines as these
will be routinely observed from galaxies during the EoR using future
observatories (e.g. JWST). LyC emission from galaxies above
$z\simeq6$, on the other hand, will not be observed due to high
neutral fraction of the IGM \citep[e.g.][]{inoue14}. This means that
calibrations based on more
readily observable features will be essential in understanding the
role of star-forming galaxies in driving reionization. 

As a first step toward achieving our goal, we selected eight galaxies at $\langle z \rangle = 3.17$, based on
30 band photometric redshift, with bright $uS$ magnitudes, indicative
of potential nonzero $f_{esc}(LyC)$. A caveat to our selection is that
at $z\sim3.17$ the $uS$ band is contaminated by Ly$\alpha$ forest
light. We opted for this redshift range in this pilot survey, however,
to minimise the effect of IGM attenuation and to maximise the
likelihood of LyC detection. We consider
galaxies selected in this way to be candidate Lyman Continuum
Emitters (LCEs). These galaxies were observed using 
MOSFIRE targeting rest-frame optical emission lines and LRIS targeting
LyC and Ly$\alpha$ emission. None of these spectral observations
showed evidence of contamination from lower redshift galaxies. Bright
emission lines from our 
MOSFIRE observations ([OII] ($\lambda\lambda$3727 {\AA}), [OIII]
($\lambda$4959 {\AA}), [OIII] ($\lambda$5007 {\AA}), and H$\beta$), as well as
Ly$\alpha$ information from LRIS, were readily detected. We estimate
$f_{esc}(LyC)$ from CLAUDS
$u$ band photometry (Sawicki et al. in preparation). These photometric
$f_{esc}(LyC)$ estimates are tentative due to the contribution to $u$
band flux from $\lambda$ $>$ 912 {\AA} photons. LyC
emission from our LRIS spectroscopy, on the other hand, has required
careful analysis, as the signal is faint, resides at short observed
wavelengths where the LRIS CCD is less sensitive, and is challenged by
ozone lines and difficulty in proper flat-fielding of the data. Improvement of
our LRIS data reduction is ongoing and resulting $f_{esc}(LyC)$ from
these observations will be the subject of future work (Me\v{s}tri\'{c} et
al. in preparation). 

Analysis of the emission line
properties of the sample provides valuable information regarding the
ionization state of the constituent galaxies. This analysis has
validated and improved our selection methodology, allowing for more
efficient selection of LCEs in subsequent observations. The results of our
emission line analysis are as follows:
\begin{itemize}
  \item The O32 ratio, often used as a proxy for ionization parameter,
    for galaxies in our sample are comparable to other LBG samples at
    a similar redshift. High O32 galaxies in our sample also overlap
    with the low O32 end of highly star-forming ``green pea'' galaxies at low redshift,
    with our highest O32 galaxy having a similar value to known
    LCEs. The majority of high redshift LAEs are found to have
    significantly higher O32 values compared to our sample. Overall,
    this suggests our sample contains galaxies with ionization
    parameters typical of  the parent population of high redshift
    star-forming galaxies but does not include extremely ionized
    members. 
  \item The ionization parameter is quantified using an iterative calculation (assuming ionization bounded
    conditions) that estimates $12+log_{10}(O/H) = 8.18-8.86$ and
    $log_{10}(q_{ion})=7.67-8.10$ for our sample, comparable to the average
    for green peas and LBGs presented in \citet{nakajima14}.
  \item 5/8 of the galaxies in our sample are also found to be
    Ly$\alpha$ emitters, and these typically have the highest
    measured values of O32 among those from our
    sample (though lower than typical O32 values measured for high
    redshift LAE samples, e.g. \citealt{nakajima16}).
  \item We find that photometric observations in the $u$-band
    (probing LyC at $z\geq2.9$) are consistent with non-zero LyC
    escape for 3/8 of the galaxies in our sample (with two additional
    galaxies having non-zero upper limits). We note that not all
    galaxies with non-zero $f_{esc}$ estimates also emit Ly$\alpha$
    photons.
  \item The two galaxies in our sample with the highest
    photometric estimates of $f_{esc}(LyC)$ have relatively low O32
    and evidence of merger activity. Shocks associated with
    merging may explain their low observed O32 values.
\end{itemize}
Difficulties in interpreting $f_{esc}(LyC)$ estimated from $u$ band
photometry has informed our selection of future samples to shift to higher redshifts. In
doing so, we expect to be able to select a cleaner sample of LCEs
as beyond $z=3.4$ the $u$ band contains only LyC photons. We note that
the selection for this pilot survey pre-dates the CLAUDS survey, thus
using this new data set allows us to more reliably select LCEs in our
on-going work (e.g. Me\v{s}tri\'{c} et al. in preparation).

In addition to the observational results presented above, we also
performed 1D photoionization modeling using the MAPPINGS V code to
predict the line ratios of H\,{\sc ii} regions with a variety of nebular
conditions. Models explore metallicities in the range 7.86 <
$12+log_{10}(O/H)$ < 8.99 and ionization parameter in the range 6.5 <
$log_{10}(q_{ion})$ < 8.5. Density bounded models allow us to explore the
additional parameter space of $f_{esc}(LyC)$ by varying the optical
depth of HI, $\tau_{HI}$. In Section \ref{section:mappings2} we
compared our $f_{esc}(LyC)$ models with the empirical relations between
O32 and $f_{esc}(LyC)$ proposed by \citet{izotov18} and \citet{faisst16} for low redshift
galaxies. The results of this analysis are: 
\begin{itemize}
  \item We have shown that the degeneracy between metallicity and
    ionization on the R23 vs O32 plane also suffers a degeneracy with
    $f_{esc}(LyC)$, a relevant issue for studies of optical line
    ratios in LCEs. Density bounded, low metallicity models indicate that, above
    $f_{esc}(LyC)=0.10$, there is little variation in O32 with
    increasing $f_{esc}(LyC)$. This may hamper the predictive power of
    this line ratio given the average uncertainties in its measurement
    at high redshift.
  \item The above point, as well as Figure \ref{fig:fesctracks},
    clearly shows that in order to reasonably estimate $f_{esc}(LyC)$
    from optical emission line ratios, an accurate estimate of galaxy
    metallicity is required. In the future, estimates of galaxy
    metallicities at $z>3-4$ will be provided using observations of
    [NII], H$\alpha$, and [SII] emission lines using JWST with the
    caveat that these values may also be affected by density bounded
    conditions in H\,{\sc ii} regions.
\end{itemize}
We stress here that photoionization models of MAPPINGS V, as well as the
commonly used photoionization code CLOUDY, typically employ homogenous spherical
or plane-parallel geometries. Observations of nearby H\,{\sc ii} regions show
extremely complex geometries implying large spatial variation in LyC
escape. Full 3D photoionization models, using e.g. the Mocassin code
\citep[e.g. Mocassin,][]{ercolano02} in the context of LyC escape
will be another interesting avenue of future research that could help
to explain peculiar observed pairings of O32 and $f_{esc}(LyC)$
(e.g. \citealp{izotov18b} galaxies with high O32 and low
$f_{esc}(LyC)$). Such simulations will also be instrumental in testing
other proposed observational signatures of LyC escape in high redshift
galaxies
\citep{steidel10,jones13,zackrisson13,zackrisson16,verhamme15,leethochawalit16,reddy16}.

The results of this paper further highlight the problems with assuming
a simple relationship between O32 and $f_{esc}(LyC)$. Other recent
works have explored this issue, focusing on the observation of
galaxies with high O32 and low (or zero) $f_{esc}(LyC)$
\citep[e.g.][]{naidu18,izotov18b}. A possible explanation for such
observations, as mentioned above, is the geometric differences between
line emission and
LyC observation due to the former being scattered into the
line-of-sight. In this work we observe a different type of
outliers with low O32 and high $f_{esc}(LyC)$, which we suggest may
result from ongoing mergers. Merger activity can both reduce O32
\citep[due to shocks, e.g.][]{epinat18} and redistribute gas and/or
induce star formation in the outskirts of the merging system
which could enhance $f_{esc}(LyC)$ \citep[see also][]{bergvall13}. Current
LCE studies may be biased against studying mergers due to selections
focusing on compact, isolated galaxies. Although this is typically done at high
redshift to avoid line-of-sight contaminants, our results highlight a
strong motivation for targeting LyC emission from mergers in future
observations. A full understanding of the relationship between O32 and
LyC with consideration of merger activity will enable the goal of
using these (or similar) nebular lines as proxies for LyC at $z>6$.

\section*{Acknowledgements}

This research was conducted by the
Australian Research Council Centre of Excellence for All Sky
Astrophysics in 3 Dimensions (ASTRO 3D), through project
number CE170100013. The authors wish to thank E. Zackrisson for useful
discussion in the early stages of this work, D. Nicholls for
assistance in constructing photoionization models using MAPPINGS V,
and the anonymous referee for their comments that have helped to
improve the focus, clarity, and relevance of this work. JC
acknowledges research support from Australian Research Council Future
Fellowship grant FT130101219. TG is grateful to the
LABEX Lyon Institute of Origins (ANR-10-LABX-0066) of the Université
de Lyon for its financial support within the programme
'Investissements d'Avenir' (ANR-11-IDEX-0007) of the French government
operated by the National Research Agency (ANR). (Some of) The data
presented herein were obtained at the W. M. Keck Observatory (programs
2015A\_W012LA, 2015A\_W033M, and 2016A\_W034LA), which is operated as a
scientific partnership among the California Institute of 
Technology, the University of California and the National Aeronautics  
and Space Administration. The Observatory was made possible by the
generous financial support of the W. M. Keck Foundation. The authors
wish to recognize and acknowledge the very significant cultural role
and reverence that the summit of Maunakea has always had within the
indigenous Hawaiian community.  We are most fortunate to have the
opportunity to conduct observations from this mountain. 

%%%%%%%%%%%%%%%%%%%%%%%%%%%%%%%%%%%%%%%%%%%%%%%%%%



\bibliographystyle{mnras}
\bibliography{refs} % if your bibtex file is called example.bib

%%%%%%%%%%%%%%%%%%%%%%%%%%%%%%%%%%%%%%%%%%%%%%%%%%

%%%%%%%%%%%%%%%%% APPENDICES %%%%%%%%%%%%%%%%%%%%%

%\appendix

%%%%%%%%%%%%%%%%%%%%%%%%%%%%%%%%%%%%%%%%%%%%%%%%%%

% Don't change these lines
\bsp	% typesetting comment
\label{lastpage}
\end{document}